\documentclass[twocolumn,epjc3]{svjour3}

\usepackage[numbers,sort&compress]{natbib}

\RequirePackage[T1]{fontenc}

\smartqed  

\RequirePackage{graphicx}
\RequirePackage[colorlinks,citecolor=blue,urlcolor=blue,linkcolor=blue]{hyperref}

\graphicspath{{figures/}} 

\usepackage{epstopdf}

\usepackage{amsmath}
\usepackage{amsfonts}
\usepackage{amssymb}
\usepackage{textcomp}
\usepackage{gensymb}
\usepackage[utf8]{inputenc}
\usepackage{hyperref}

\usepackage{physics}
\usepackage{slashed}
\usepackage{simplewick}

\usepackage{feynmp-auto}

\newcommand{\diff}{\textrm{d}}

\journalname{Eur. Phys. J. A}

\begin{document}

\title{Progress and Opportunities in Backward angle ($u$-channel) Physics}

\author{C.~Ayerbe~Gayoso\thanksref{addr12}
        \and Ł.~Bibrzycki\thanksref{addr15}
        \and S.~Diehl\thanksref{addr1,addr1a}
        \and S. Heppelmann\thanksref{addr13}
        \and D.W.~Higinbotham\thanksref{addr10}
        \and G.M.~Huber\thanksref{addr2}
        \and S.J.D.~Kay\thanksref{addr2}
        \and S.R.~Klein\thanksref{addr3}
        \and J.M.~Laget\thanksref{addr10}
        \and W.B.~Li\thanksref{e4,addr4, addr10}
        \and V.~Mathieu\thanksref{ucm,ub}
        \and K.~Park\thanksref{addr5}
        \and R.J. Perry\thanksref{addr14}
        \and B.~Pire\thanksref{addr6}
        \and K.~Semenov-Tian-Shansky\thanksref{addr7,addr7b}
        \and A.~Stanek\thanksref{addr3}
        \and J.R.~Stevens\thanksref{addr4}
        \and L.~Szymanowski\thanksref{addr8} 
        \and C.~Weiss\thanksref{addr10} 
        \and B.-G.~Yu\thanksref{addr11} 
}

\thankstext{e4}{Corresponding Author E-mail: billlee@jlab.org}

\institute{Mississippi State University, Starkville, MS 39762, USA\label{addr12}
   \and
          Institute of Computer Science, Pedagogical University of Krakow, 30-084 Krak\'ow, Poland
          \label{addr15}
          \and
          Justus Liebig Universit\"at Gie{\ss}en, 35390 Gie{\ss}en, Germany \label{addr1}
          \and
          University of Connecticut, Storrs, CT 06269, USA \label{addr1a}
          \and
          University of California, Davis, CA 95616, USA\label{addr13}
          \and 
          Thomas Jefferson National Accelerator Facility, Newport News, VA 23606, USA\label{addr10}
          \and
          University of Regina, Regina SK  S4S~0A2 Canada \label{addr2}
          \and
          Lawrence Berkeley National Laboratory, Berkeley, CA 94720, USA \label{addr3}
          \and
          William \& Mary, Williamsburg VA 23185, USA\label{addr4}
          \and
          Departamento de F\'isica Te\'orica, Universidad Complutense de Madrid and IPARCOS, 28040 Madrid, Spain\label{ucm}
          \and
          Departament de F\'isica Qu\`antica i Astrof\'isica and Institut de Ci\`encies del Cosmos, Universitat de Barcelona, E08028, Spain \label{ub}
          \and
          Hampton University Proton Therapy Institute, 40 Enterprise Parkway, Hampton, VA 23666 \label{addr5}
                \and
          National Yang Ming Chiao Tung University, Hsinchu 30010, Taiwan\label{addr14}
          \and
          CPHT, CNRS, Ecole polytechnique, I.P. Paris, 91128-Palaiseau, France \label{addr6}
          \and
          National Research Centre Kurchatov Institute: Petersburg Nuclear Physics Institute, 188300 Gatchina, Russia \label{addr7}
          \and
          Higher School of Economics, National Research University, 194100 St. Petersburg, Russia \label{addr7b}
          \and
          National Centre for Nuclear Research, NCBJ, 02-093 Warsaw, Poland \label{addr8}
          \and
          Korea Aerospace University, Goyang 10540, South Korea\label{addr11}
}

\date{\today}

\maketitle

\begin{abstract}
%
 Backward angle ($u$-channel) scattering provides complementary information for studies of hadron spectroscopy and structure, but has been less comprehensively studied than the corresponding forward angle case. As a result, the physics of $u$-channel scattering poses a range of new experimental and theoretical opportunities and questions. We summarize recent progress in measuring and understanding high energy reactions with baryon charge exchange in the $u$-channel, as discussed in the first {\em Backward angle ($u$-channel) Physics Workshop}. In particular, we discuss backward angle measurements and their theoretical description via both hadronic models and the collinear factorization approach, and discuss planned future measurements of $u$-channel physics. Finally, we propose outstanding questions and challenges for $u$-channel physics.
\end{abstract}

\section{Introduction}

The~\href{https://www.jlab.org/conference/BACKANGLE}{first workshop on Backward angle ($u$-channel) physics} was held virtually in September 2020. Twenty presentations were given over three days, to more than 50 registered participants.  This white paper summarizes the existing measurements, theoretical developments and opportunities presented and discussed at this workshop.

\subsection{Physics motivation}
\label{sec:motivation}

Studying nucleon structure through probes of real and virtual photons has been a key objective of the hadron physics community. 
%
%
%
%
%
In this document, we focus on one specific type of interaction, known as the backward angle (or $u$-channel) exclusive interaction, where the squared momentum transfer $u$ between a produced meson $M$ and the target is such that
\begin{equation}
\vert u \vert =\vert (p_{M}-p_{N})^2 \vert \ll \vert t \vert =\vert (p'_{N}-p_{N})^2 \vert.
\end{equation}

\noindent The non-intuitive nature of such an interaction is revealed in the following example: a real ($\gamma$) or virtual photon probe ($\gamma^*$) is induced (by the accelerated electron) and interacts with the proton target at rest. The recoiling nucleon absorbs most of the momentum transfer from the probe and travels forward, whereas the produced meson remains close to the target, nearly at rest. This type of reaction is sometimes referred to as a ``knocking a proton out of a proton'' process, and offers improved access to the valence quark plus sea components of the nucleon wave function. To ensure the final state dynamics are not dominated by resonance contributions, the invariant mass $W=\sqrt{(p_{\gamma}+p_{N})^2}$ is chosen to be above nucleon resonance region ($W>2.0$~GeV). In addition, all final state particles must be directly detected or indirectly reconstructed (using the missing mass reconstruction technique) to ensure exclusivity.


The exclusive $u$-channel interaction in the deep regime (i.e. at large $Q^2=-(p_e-p_{e'})^2$) is unexplored territory. Limited past photoproduction measurements which led to predictions of a small (1/100 of the $t$-channel peak) rise in cross section in such a kinematic regime discouraged most experimental efforts.
However, recent experimental results on exclusive meson electroproduction indicate a much larger observed cross section (1/10 of the $t$-channel peak)~\cite{wenliang17}.
Due to limited available data, developing a unified description of $u$-channel interactions remains an outstanding challenge. The objectives of the {\em Backward angle (u-channel) Physics Workshop} were to summarize the current theoretical and experimental developments, to establish a consensus on the priorities in terms of data taking and theory developments, and to derive a coherent strategy going forward. 

The $u$-channel physics strategy is intended to answer the following questions:
\begin{itemize}
\item{Do the  backward-angle observables for meson electroproduction exhibit a transition from a soft\\(hadronic) to a hard (partonic) regime?}
\item{Do other channels (for instance DVCS or TCS) indicate a similar transition?}
\item{Is the cross section scaling behavior in $Q^2$ predicted by the collinear QCD factorization scheme observable at medium energies?}
\item{Do polarization observables provide a clearer understanding of the underlying dynamics?}
\end{itemize}


The answers to these questions are critically important in helping to develop a unified Regge model to validate our understanding of the relevant exchange mechanisms and to probe the soft-hard transition of QCD in different kinematic regions. These insights may point to unique nucleon structure information which may only be manifest in backward-angle observables, complementary to their forward-angle counterparts.

\subsection{Outline of the Paper}


The analysis efforts of the JLab 6 GeV program have shown evidence for the existence of a backward-angle ($u$-channel) peak in multiple exclusive meson electroproduction channels above the resonance region ($W>2$~GeV). In 2018, the CLAS collaboration reported \cite{Park:2017irz} the first cross section measurement on exclusive $\pi^+$ electroproduction off the proton in near-backward kinematics. In 2019, a study from Hall C \cite{Li:2019xyp} revealed the existence of a backward-angle peak in the exclusive $\omega$ electroproduction cross section. Most interestingly, the separated cross section ratio at $Q^2=2.2$~GeV$^2$ indicated a dominance of the transverse over longitudinal cross section. Furthermore, the recently published Beam Spin Asymmetry analysis \cite{Diehl:2020uja} of exclusive $\pi^+$ meson production by the CLAS collaboration has shown evidence of a sign change in the interference contribution ($\sigma_{LT}$) as the interaction changes from $-t_{min}$ to $-t_{max}$, or forward angle to backward angle regimes. These results are documented in Sec.~\ref{sec:current_results}. 


In terms of the theoretical efforts in describing $u$-channel physics interactions, there are two types of models that offer different types of exchange mechanisms. The first is a hadronic Regge-based model, which explores the meson-nucleon dynamics of the hadron production reaction.  The other is a GPD-like model, known as the Transition Distribution Amplitude (TDA), which similarly involves hard-soft factorization and universal nonperturbative objects, in this case describing the baryon-to-meson transition.  Here, it is important to note both types of models have met some success in describing existing data at JLab 6~GeV kinematics. Descriptions of both models are documented in Secs.~\ref{sec:theory}.



Sec.~\ref{sec:next_step} reflects a vision of the future perspective on backward angle physics shared by the workshop participants. Sec.~\ref{sec:soft_hard_transition}-\ref{sec:forward-backward_asymmetry} outlines the feasible approaches and measurements that are considered necessary for advancing the theory development in both Regge based in subsection \ref{sec:$u$-channel regge_prospective} and in TDA models, emphasizing the nucleon to photon TDAs needed for the description of backward DVCS and TCS, in subsection \ref{sec:btcs}. 



Looking ahead, $u$-channel interactions will be studied at multiple next generation experimental facilities. The upcoming experiments at Jefferson Lab Hall C (described in  Sec.~\ref{sec:hallc_12gev}), E12-20-007 \cite{E12-20-007}, E12-19-006 \cite{E12-19-006}, E12-09-011 \cite{E12-09-011}, will measure differential cross sections  (with the possibility of  longitudinal/transverse separation) for $\pi^0$, $\omega$ and $\phi$ mesons in the extreme $u$-channel kinematics ($u\rightarrow u_{min}$) up to at least $Q^2=6$ GeV$^2$. Directly inspired by these questions, one could utilize the synergy between exclusive $\omega$ photoproduction (as an example) measured at GlueX and electroproduction measured (at $Q^2<2$~GeV$^2$) at CLAS12. Note that both are state of the art detector packages with large acceptance. This is further elaborated in Sec.~\ref{sec:GlueX_CLAS12}. Section~\ref{sec:EIC} outlines the critical role an Electron-Ion Collider \cite{AbdulKhalek:2021gbh, Anderle:2021wcy} will play in probing exclusive $u$-channel meson production. Subsection \ref{sec:UPC} discusses   new opportunities opened by ultra peripheral collisions in the backward-angle regime. 
Finally, Sec.~\ref{sec:panda} describes feasibility studies  \cite{Singh:2014pfv,Singh:2016qjg} for exclusive $\pi^0$ production through the $p\overline{p}$ annihilation processes, $\overline{p} + p \rightarrow \gamma^* + \pi^0$ and  $\overline{p} + p \rightarrow J/\psi + \pi^0$,  at $\overline{\textrm{P}}$ANDA/FAIR. In these annihilation reactions, both forward and backward regions correspond to a baryonic exchange.


\section{Recent Experimental Results from JLab 6 GeV}
\label{sec:current_results}
For the reactions discussed here, the cross-section for the electroproduction of a meson $M$:
\begin{equation}
 e(k,h) + N(p_N) \to e'(k')+ M(p_M)+ N'(p'_{N})   
\end{equation}
is expressed in terms of $\sigma_L, \sigma_T, \sigma_{LT}, \sigma_{LT'}, \sigma_{TT}$ through the usual equations~\cite{Pire:2021hbl,Arens:1996xw}:
\begin{eqnarray}
&&
\frac{d^4 \sigma (e N \to e'N'M)}{ds dQ^2 d \phi dt}= \frac{\alpha_{\rm em} (s-m_N^2)}{4 (2 \pi)^2 ( {k}_0^{\rm LAB})^2 m_N^2 Q^2 (1-\varepsilon)} \times
\nonumber \\ &&
\left[
{
\frac{d \sigma_T}{dt}
+ \varepsilon \frac{d \sigma_L}{dt}+
\varepsilon \cos 2\phi \frac{d \sigma_{TT}}{dt}+
\sqrt{2 \varepsilon (1+\varepsilon)} \cos \phi \frac{d \sigma_{LT}}{dt} }\right.
\nonumber \\
&& 
+ h[2\varepsilon(1-\varepsilon)]^{1/2}\frac{d\sigma_{LT'}}{dt} \sin\phi  - h(1-\varepsilon)^{1/2}\frac{d\sigma_{TT'}}{dt} \left. \right].
\label{Def_CS_Kroll}
\end{eqnarray}
Here
$s=(p_N+q)^2 \equiv W^2$,
$t=(p'_N-p_N)^2$ and $Q^2=-q^2$
are the usual Lorentz invariants,
$\phi$
is the angle between the leptonic and hadronic planes;
and
${k}_0^{LAB} $
is the initial state electron energy in the laboratory (LAB) frame  (electron beam energy).
$\varepsilon$
is the polarization parameter of the virtual photon
that expresses the ratio of longitudinal to transverse photon flux,
 $x_B=\frac{Q^2}{2 p_N \cdot q}$ and $y=\frac{p_N \cdot q}{p_N \cdot k}$
are the usual dimensionless variables; 
${k'}_0^{LAB} $
is the energy of the final state electron in the LAB frame and
$\theta^{LAB}_e $
is the electron scattering angle in the LAB frame; $h$ is the incoming electron helicity. 

\subsection{$\pi^+$ electroproduction with CLAS 6}

\subsubsection{Hard exclusive $\pi^{+}$ electroproduction cross section in backward kinematics
\label{sec:parkdata}}

\begin{figure} 
\centering
  \includegraphics[width=0.47\textwidth]{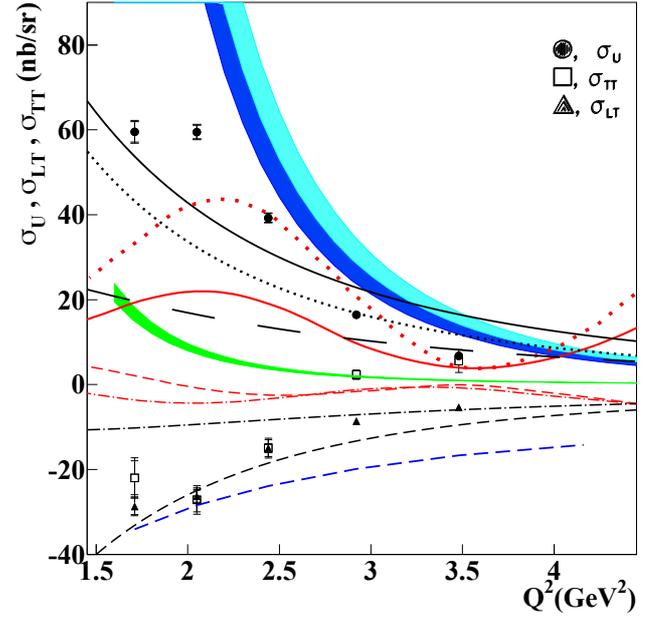}
  \caption{The structure functions $\sigma_U = \sigma_T+\varepsilon \sigma_L$ (solid dot), $\sigma_{TT}$ (square) and $\sigma_{LT}$ (triangle) as a function of $Q^2$. The inner error bars are statistical and the outer error bars are the combined systematic and statistical uncertainties in quadrature.  The bands refer to model calculations of $\sigma_U$ in the TDA description, green band: BLW NNLO, dark blue band: COZ, and light blue band: KS (see \cite{Park:2017irz} and refs. therein for the meaning of these models). The lower blue short-dashed line represents an educated guess to fit the higher twist cross sections $\sigma_{LT}$ and $\sigma_{TT}$ in the TDA picture. The red curves are the ``Regge'' predictions (by JML18) of \cite{guidal97, Laget:2019} for solid: $\sigma_U$, dashed curve: $\sigma_{LT}$, dot-dashed: $\sigma_{TT}$. An updated $\sigma_U$ calculation from JML18 model ~\cite{Laget:2021} are shown in red dotted curve. Regge calculations which consider parton contributions (described in Sec.~\ref{sec:Regge}) to $\sigma_{U}$, $\sigma_{T}$, $\sigma_{L}$, $\sigma_{TT}$ and $\sigma_{LT}$ are shown in black solid, black dotted, black  long-dashed, black dot-dashed and black short-dashed, respectively. }
  \label{fig:CLAS_cross_section_u_channel} 
\end{figure}


The cross section for hard exclusive $\pi^{+}$ production ($e p \rightarrow e' n \pi^+$) from a polarized electron beam interacting with an unpolarized hydrogen target has been studied with the CLAS detector in the backward kinematic regime in Ref. \cite{Park:2017irz}. 

Fig.~ \ref{fig:CLAS_cross_section_u_channel} shows the $Q^2$-dependence of $\sigma_U=\sigma_T+\varepsilon\sigma_L$, $\sigma_{LT}$ and $\sigma_{TT}$, obtained at the average kinematics of $W=2.2$ GeV and $-u=0.5$ GeV$^2$. All three cross sections have a strong $Q^2$-dependence. The TDA formalism predicts that the transverse amplitude dominates 
at large $Q^2$. In order to validate the TDA approach, it is necessary
to separate $\sigma_T$ from $\sigma_L$ and check that $\sigma_T\gg  \sigma_L, \sigma_{TT}$ and $\sigma_{LT}$.
With only this set of data at fixed beam energy, the CLAS detector cannot experimentally separate $\sigma_T$ and $\sigma_L$. After examining the angular dependency of the $sigma_U$, the result shown $\sigma_{TT}$ and $\sigma_{LT}$ are roughly equal in magnitude and have a
similar $Q^2$-dependence. Their significant size (about 50\% of $\sigma_U$) imply an important contribution of the transverse amplitude in the cross section.  Furthermore, above $Q^2=2.5$ GeV$^2$, the trend of $\sigma_U$ is qualitatively consistent with the TDA calculation, yielding the characteristic $1/Q^8$ dependence expected when the backward collinear factorization scheme is approached.



In a ``Regge'' based framework (to be introduced in Sec.~\ref{sec:Regge}), the Born terms in the Reggeized amplitude are consisting of neutron and $\Delta^0$ exchanges in the $u$-channel which are gauge-invariant with the magnetic coupling to virtual photon in both exchanges.

The first version (red curves) uses point like meson-baryon vertices and canonical Regge propagators ~\cite{guidal97,Laget:2019} (solid line for $\sigma_U$, dashed line for $\sigma_{LT}$ and dot-dashed for $\sigma_{TT}$), and includes the contribution of $u$-channel cuts~\cite{Laget:2021} (dotted line for $\sigma_U$). We refer the reader to \cite{Laget:2021} for details. The second version (black curves) is presented in Sec. \ref{sec:$u$-channel regge_prospective} and relies on a partonic description of the meson-baryon vertices. The $Q^2$-dependence of $\sigma_U$, $\sigma_T$, $\sigma_L$, $\sigma_{LT}$ and $\sigma_{TT}$ are presented at $W=2.2$ GeV, $u=-0.5$ GeV$^2$ and $\epsilon=0.5$ to show that the $\sigma_U$ is comparable with that of the TDA predictions. The transverse cross section $\sigma_T$ is larger than any other components of cross sections, which supports the validity of the TDA approach to access the $u$-channel physics at backward angles.  



\subsubsection{Hard exclusive $\pi^{+}$ electroproduction beam spin asymmetry in a wide range of kinematics}

\begin{figure} 
\hspace{0mm}
  \includegraphics[width=0.47\textwidth]{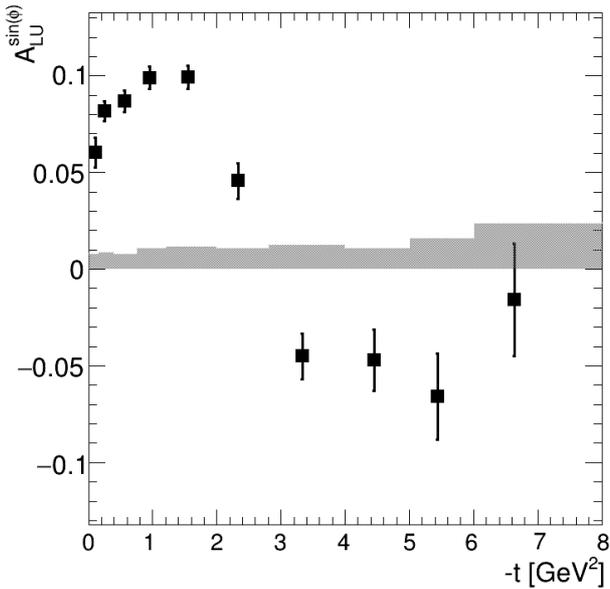}
\caption{$A_{LU}^{\sin\phi}$ as function of $-t$ measured with CLAS in the deep inelastic regime (W $>$ 2 GeV, $Q^{2} >$ 1 GeV$^2$). The maximal accessible value of $-t$ is $\approx 8.8$ GeV$^{2}$. The shaded area represents the systematic uncertainty. These results were first published in Ref.\cite{Diehl:2020uja}.}
\label{fig:CLAS_t_ALU_PRL} 
\end{figure}

Based on the beam spin asymmetry, the $A_{LU}^{\sin\phi}$ moment has been extracted. $A_{LU}^{\sin\phi}$ is proportional to the polarized structure function $\sigma_{LT^\prime}$,
\begin{equation}\label{eq:ALU}
\begin{aligned}
\hspace{6em} 	A_{LU}^{\sin\phi} = \frac{\sqrt{2 \varepsilon (1 - \varepsilon)}~\sigma_{LT^{\prime}}}{\sigma_{T} + \varepsilon \sigma_{L}}
\end{aligned}
\end{equation}
where the structure functions $\sigma_{L}$ and $\sigma_{T}$ correspond to longitudinal and transverse virtual photons, and $\varepsilon$ describes the ratio of their fluxes.
Due to the large acceptance of CLAS, it was possible to map out the full kinematic region in $-t$ from very forward kinematics ($-t/Q^{2} \ll 1$) where a description based on Generalized Parton Distributions (GPD) can be applied, up to very backward kinematics ($-u/Q^{2} \ll 1$, $-t$ large)
where a description based on baryon-to-meson Transition Distribution Amplitudes (TDA) is assumed to be valid. The result is shown in Fig. \ref{fig:CLAS_t_ALU_PRL}.

A clear transition from positive values of $A_{LU}^{\sin\phi}$ in the forward regime to negative values in the backward regime can be observed, with the sign change occurring near 90$^{\circ}$ in the center-of-mass \cite{Diehl:2020uja}. It was found that this sign change between the forward and backward kinematic regime is independent of $Q^2$ and $x_B$ within the kinematics accessible with CLAS \cite{Diehl:2020uja}. By performing accurate measurements over a wide range of $Q^{2}$, $x_{B}$ and $-t$, CLAS can explore the transition from hadronic to partonic reaction mechanisms \cite{Diehl:2020uja}.

\subsection{Hall C 6 GeV $u$-channel $\omega$ Production
\label{sec:hallc_omega}}


A study from JLab Hall C, of backward-angle $\omega$ cross sections
from exclusive electroproduction $e p \rightarrow e^{\prime} p \omega$, was
published in 2019 \cite{Li:2019xyp}.  The scattered electron and forward-going
proton were detected in the HMS and SOS high precision spectrometers, and the
low momentum rearward-going $\omega$ was reconstructed using the missing mass
reconstruction technique, described in~\ref{app:missmass}. Since the missing mass reconstruction method does
not require the detection of the produced meson, this allows the analysis  to
cover a kinematic range inaccessible through direct-detection experiments.

The analyzed data were part of experiment E01-004 (F$_{\pi}$-2) \cite{horn06,
 blok08}. The primary objective of the experiment was to detect coincidence
$e$-$\pi$ at forward-angle, where the backward-angle $\omega$ ($e$-$p$
coincidence) events were fortuitously acquired in the same data set.  Data were
acquired at $Q^2=1.60$ and 2.45 GeV$^2$,  at $W=2.21$ GeV (above the resonance region), with $-u\sim -u_{\rm min}$; $-t\sim -t_{\rm max}$. 

\begin{figure}[htb]
\centering
\includegraphics[scale=0.43]{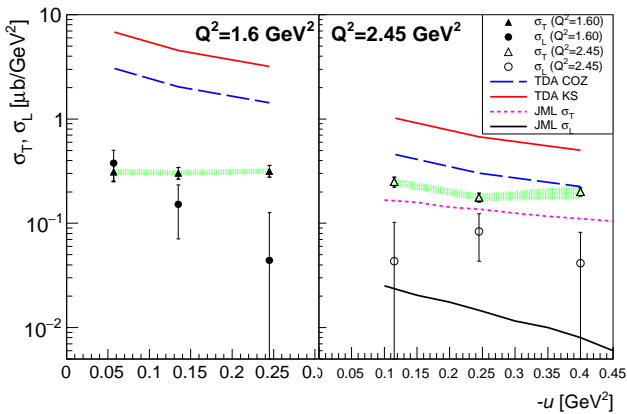}
\caption{$\sigma_{ T}$ (triangles), $\sigma_{ L}$ (squares) as function
  of $-u$, at $Q^2=1.6$ GeV$^2$ (left), 2.45 GeV$^2$ (right). 
   For the lowest $-u$ bin, $u^{\prime} = u- u_{\rm min}\approx0$. 
   TDA predictions for $\sigma_{ T}$: COZ~\cite{chernyak89} (blue dashed
   lines) and KS~\cite{king87} (red solid lines). The predictions were calculated at the specific $\overline{Q^2}$, $\overline{W}$ values of each $u$ bin. The predictions at three $u$ bins are joined by straight lines for visualization purpose.  The green bands indicate correlated systematic uncertainties for $\sigma_{ T}$, the uncertainties for $\sigma_{ L}$ have similar magnitudes. The JML18 model \cite{Laget:2021} reproduces $\sigma_L$ and $\sigma_T$ at $Q^2=$2.45 GeV$^2$, as shown by the black solid and dotted magenta curves, respectively. The full calculation includes $\sigma_{LT}$ and $\sigma_{TT}$ as presented in \cite{Laget:2021}. \label{fig:sigt}}
\end{figure}

The extracted $\sigma_{\rm L}$ and $\sigma_{\rm T}$ as a function of $-u$ at
$Q^2=1.6$ and 2.45~GeV$^2$ are shown in Fig.~\ref{fig:sigt}.  The two sets of
TDA predictions for $\sigma_{\rm T}$ each assume different nucleon DAs as
input.  From the general trend, the TDA model offers a good description of the
falling $\sigma_{\rm T}$ as a function of $-u$ at both $Q^2$ settings.  This is
very similar to the backward-angle $\pi^+$ data from CLAS in Sec.~\ref{sec:parkdata}.
Together, the data sets are suggestive of early TDA scaling.
The behavior of $\sigma_{\rm L}$  differs greatly at the two $Q^2$
settings. At $Q^2=$1.6~GeV$^2$, $\sigma_{\rm L}$ falls almost exponentially as
a function of $-u$; at $Q^2=$2.45~GeV$^2$, $\sigma_{\rm L}$ is constant
near zero (within one standard deviation). Note that  the TDA model  predicts a small - higher twist - $\sigma_{\rm L}$ contribution, which falls faster with $Q^2$ than the leading twist $\sigma_{\rm T}$ contribution.

\begin{figure}[htb]
\centering
\includegraphics[scale=0.44]{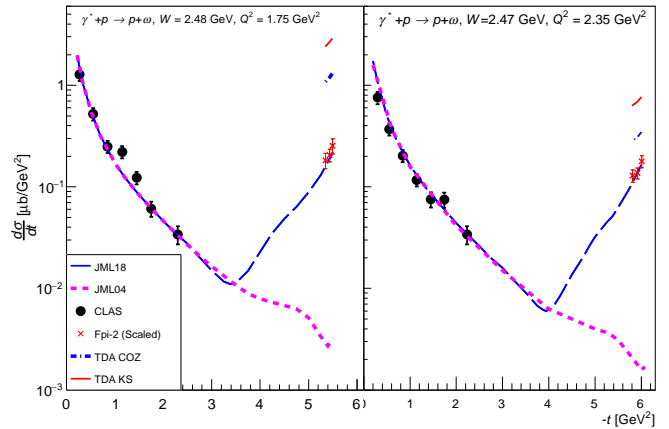}
\caption{Exclusive $\omega$ electroproduction cross section as a function $-t$
  at $Q^2=1.75$ (left panel) and $Q^2=2.35$~GeV$^2$ (right panel). The CLAS
  data are the black dots in the near-forward kinematics region
  ($-t<2.5$~GeV$^2$), and the Hall C are the red crosses in the backward
  region ($-t>5$~GeV$^2$), $W$-scaled (from 2.2 to 2.46 GeV) to the kinematics of the CLAS data.  The blue and magenta dashed thick lines are
  Regge trajectory based JML04 and JML18 predictions, respectively. The
  short curves above the Hall C data are TDA predictions based on
  COZ~\cite{chernyak89} (blue solid) and KS~\cite{king87} (red solid) DAs.
\label{fig:omega_t_slope}
}
\end{figure}

Combined data from CLAS \cite{Morand05} and Hall C cover both forward and
backward-angle kinematics, and jointly form a complete $t$ evolution picture of
the $p(e,e'p)\omega$ reaction.  The Hall C data are ``gently'' scaled to those
of the CLAS data, $W\sim2.48$~GeV$^2$, $Q^2=1.75$ and 2.35 GeV$^2$.  In
addition to this scaling, the Hall C $u$ dependent cross sections are
translated to the $t$ space of the CLAS data.  Fig.~\ref{fig:omega_t_slope}
indicates strong evidence of the existence of the backward-angle peak at
$-t>5$~GeV$^2$ for both $Q^2$ settings, with strength $\sim$1/10 of the
forward-angle cross section. Previously, the ``forward-backward peak''
phenomenon was only observed in $\pi^+$ photoproduction data~\cite{
  guidal97, anderson69, anderson69b, boyarski68}, which was successfully
interpreted using a Regge trajectory based  model~\cite{guidal97}.
\begin{figure}[ht]
\centering
\includegraphics[scale=0.35]{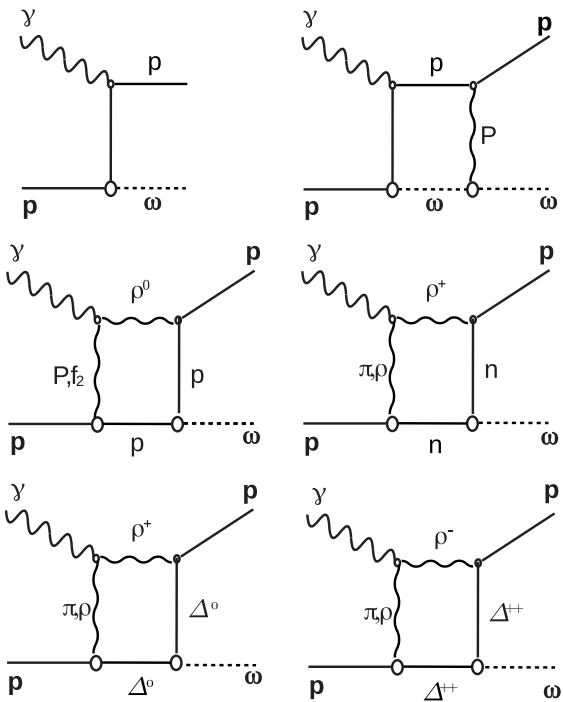}
\caption{Regge exchange diagrams for the $p(e,e'p)\omega$ reaction at backward angles  in the JML18 model. Top: nucleon Regge pole exchange (left) and elastic re-scattering (right).
Middle: $\rho^0p$ (left) and $\rho^{+}n$ (right) inelastic scattering.  Bottom: $\rho^+\Delta^0$ and $\rho^-\Delta^{++}$ inelastic scattering.  For more details, see Ref.~\cite{Laget:2021}. 
\label{fig:omega_u_cut}
}
\end{figure}

The Regge based  model \cite{guidal97} provides a natural description of JLab $\pi$
electroproduction cross sections over a wide kinematic range without destroying
good agreement at $Q^2$=0 \cite{laget10,laget11}.  Two Regge-based 
predictions are plotted in Fig.~\ref{fig:omega_t_slope}: JML04~\cite{laget04}
(prior to F$_\pi$-2 data) and JML18~\cite{Laget:2021}. JML04 includes the near-forward Regge
contribution at $-t<1$~GeV$^2$ and $N$-exchange in the $u$-channel with a
$t$-dependent cutoff mass, it significantly under predicts the backward-angle
cross section.  JML18 includes in addition $\rho N$ and $\rho\Delta$
rescattering inside the nucleon (Regge Cuts), see Fig.~\ref{fig:omega_u_cut}.  As
shown, JML18 offers an excellent description of the combined data within a
consistent framework. In particular, the $-u$ dependence and the strength of
the backward angle peak is described well at both $Q^2$ settings.  The
inelastic exchange diagrams (Fig.~\ref{fig:omega_u_cut} bottom panels) are the main
contributors to the observed backward-angle peak, with one third of the
contribution coming from the $\rho^0$-$\omega$ transition (bottom left), and
the rest coming from $\rho^+$-$N$ and $\Delta$ resonance (bottom right). The model reproduces also  $\sigma_L$ and $\sigma_T$ separately (Fig.~\ref{fig:sigt}).


To summarize, the Hall~C result accessed a previously
ignored kinematic domain, which can directly challenge the $\sigma_{\rm T}$
dominance prediction ($\sigma_{\rm T} \gg \sigma_{\rm L}$) by TDA in
backward-angle exclusive meson electroproduction, as well as with Regge approach. In combination with a large acceptance detector, such as CLAS-12, one could systematically study the complete $t$ evolution of a given interaction, thus unveiling new aspects of nucleon structure.  The separated cross sections show indications of the $\sigma_{\rm T} \gg \sigma_{\rm L}$ for $ep\rightarrow e^{\prime}p\omega$, qualitatively consistent with the TDA factorization approach in backward-angle kinematics. Further $u$-channel data  from
JLab Hall C (Sec.~\ref{sec:hallc_12gev}), will allow the $\sigma_{\rm T} \gg
\sigma_{\rm L}$ and $1/Q^8$ scaling predictions to be checked with much greater
authority, as well as to address the issue of the transition between Regge and Parton descriptions.

\section{Theoretical landscape}
\label{sec:theory}

\subsection{Regge model for $\omega$-meson electroproduction at backward angles}
\label{sec:Regge}
The Regge description of mesons photo- and electroproduction has already been summarized in \cite{Laget:2019} and updated at backward angles in~\cite{Laget:2021} which we refer the reader to for details. In this section we present a concurrent approach.



%
The Regge theory-based description is complementary to the partonic description and
does not exclude the possibility to obtain a description within a collinear factorization 
framework at certain kinematical range.   
Moreover, it can be employed to specify the kinematical regime in which the reaction mechanism 
is subject to a change, which may signal the applicability of  partonic interpretation.

The reaction cross sections exhibit a distinct
difference between forward and backward processes. The peak of the
differential cross section at backward angles is, in general, smaller than that of the forward cross section by one order of magnitude, and is mainly due to the exchange of nucleon and $\Delta$ resonances in the $u$-channel.
In the process where the scattering angle $\theta$ approaches the forward or backward limit, the impact parameter $b$ converges from a large value to a smaller one. Therefore, it is natural to consider the
substructure via the form factor between hadron interactions in
the latter process, because the backward process can be seen as
the reaction that creates a spatial overlap between target and
produced particles in an intuitive picture. The internal structure
of the hadron, expressed in terms of partonic degrees of
freedom, can be embodied in this form factor~\cite{guidal}, e.g.,
the strong form factor in the $u$-channel baryon exchange in the
Reggeized model~\cite{yu}.

In $\pi$ electroproduction, the transverse cross section
$\sigma_T$ is found to be larger than expected and difficult to reproduce in
hadronic models, whereas the longitudinal cross section $\sigma_L$ is
well described by $t$-channel meson exchange. 
In Regge models in particular, such a
difficulty in $\sigma_T$ was claimed to be resolved by enhancing the contribution of nucleon exchange
with a complicated role of the nucleon electromagnetic form factors~\cite{kaskulov:2010}. Alternatively, it was found that a large cutoff
mass $\Lambda_p=1.55$ GeV (defined in Eqn.~\ref{eqn:dipoleff}) provides a good description of the data ~\cite{tChoi:2015}.

This is similar to backward $\omega$ electroproduction, where 
the large $\sigma_T$ is in question. Since $\omega$ electroproduction
allows only nucleon exchange in the $u$-channel, a deep dip of the nucleon
trajectory is expected. However, the dip is 
assumed to be filled with partonic contributions from the measured cross section~\cite{yu}. 
Then, the issue at hand is how to describe the backward cross section measured 
in Hall~C with the possibility of partonic contributions.
In the $u$-channel Regge model for electroproduction, it is better to treat nucleon electromagnetic form factors as the dipole form factor of photon momentum squared $Q^2$, so that a consistency with the photoproduction case can be recovered as\footnote{The nucleon electromagnetic form factors parameterized in terms of parton distribution functions, as in Ref.~\cite{guidal} can, in practice, be obtained equally by adjusting the cutoff mass $\Lambda$ in the corresponding dipole form factor.} $Q^2\rightarrow 0$.  The partonic contribution can be implemented in the meson-baryon form factor at the $VNN$ vertex in the $u$-channel via a parameterization of the parton distribution function (PDF).  Therefore, the Feynman diagram for the nucleon exchange in the $u$-channel is similar to the transverse distribution amplitude (TDA)~\cite{Pire:2015kxa} (see right panel of Fig.~\ref{fig:bTCS} in Sec~\ref{sec:btcs}).

For the Reggeized nucleon exchange in the $u$-channel, we utilize an extended version of Ref.~\cite{yu} to electroproduction, which consists of hadronic and partonic parts, with the relative phase adjusted by the angle  $\phi(u)$, i.e.~\cite{yu},
\begin{eqnarray}
&&\mathcal{M}=(M_s+M_u)_{Born} \times (u-m_N^2)\nonumber \\&&
   \hspace{4em} \times \biggl(\mathcal{R}^N(s,u)+e^{i\phi(u)}F^{(s)}(u)\widetilde{\mathcal{R}}^N(s,u)\biggr).
    \label{eqn:nucleon2}
\end{eqnarray}
Here, the $u$-channel proton exchange needs the $s$-channel proton pole for gauge invariance of the Born amplitude.  A modified nucleon Reggeon $\widetilde{\cal R}^N$ is assumed further in the presence of parton contributions. Hereafter, we will regard the Reggeized baryon exchange amplitude in Eqn.~(\ref{eqn:nucleon2}) as an extended framework that includes some partonic contribution to backward hadron reactions.  For the conventional approach, we simply do without the second term, i.e. the partonic part in Eqn.~(\ref{eqn:nucleon2}).

The Regge propagator of a spin-$J$ baryon is
\begin{eqnarray}
&&\mathcal{R}^J(s,u)=\alpha_J' \Gamma[J-\alpha_J(u)] \frac{1}{2} \biggl[
1+\eta e^{i\pi(\alpha_J(u)-0.5)}\biggr]\nonumber \\&&\hspace{1.75cm}
\times \biggl( \frac{s}{s_0} \biggr)^{\alpha_J(u)-J},
    \label{eqn:regge}
\end{eqnarray}
with $s_0=1$~GeV$^2$ and $J=1/2$ and $3/2$ for nucleon and
$\Delta$, respectively. The signature $\eta=+1$ is chosen for the
nucleon and $\eta=-1$ for the non-degenerate $\Delta$ trajectory.
The nucleon trajectory is fixed to be $\alpha_N(u)=0.9u-0.365$ in
order to reproduce the observed dip in the cross section at $u=-0.15$
GeV$^2$ by the nonsense-wrong signature zero from the vanishing of
the canonical phase in Eqn. (\ref{eqn:regge}).

Employing the nucleon Born terms in Eqn.~(\ref{eqn:nucleon2}), as
given in Ref.~\cite{yu}, the Gross-Riska prescription is adopted
for gauge invariance of the electromagnetic form factors at the
$\gamma^*NN$ vertex \cite{gr},
\begin{eqnarray}
\label{eqn:vertex}
\hspace{2em}\Gamma_{\gamma^*
NN}(k)=e\left(\widetilde{F}^N_1(k^2)\rlap{/}\epsilon-
\frac{F_2^N(k^2)}{4m_N}\left[\rlap{/}\epsilon,\,\rlap{/}k\right]\right)\label{vertex1}
\end{eqnarray}
with
\begin{eqnarray}
\hspace{2em}&&{\widetilde
F}^N_1(k^2)\rlap{/}\epsilon=\left[F^N_1(k^2)-F^N_1(0)\right]
\left(\rlap{/}\epsilon-\rlap{/}k {\epsilon\cdot \frac{k}{k^2}}\right)
\nonumber\\&&\hspace{4em}
+F^N_1(0)\rlap{/}\epsilon\,,
\end{eqnarray}
where $\epsilon$ stands for the photon polarization vector.
The nucleon electromagnetic form factors are
\begin{eqnarray}
\label{eqn:dipoleff}
\hspace{4em}&&F_1^p(Q^2)=\left(\frac{1+\tau\mu_p}{1+\tau}\right)
\left(1+\frac{Q^2}{\Lambda^2_p}\right)^{-2},
\nonumber\\
\hspace{4em}&&F_2^p(Q^2)=\left(\frac{\kappa_p}{1+\tau}\right)
\left(1+  \frac{Q^2}{\Lambda^2_p}\right)^{-2}\hspace{0.3cm},
\end{eqnarray}
with $\tau=Q^2/4m_N^2$ and $\mu_p=1+\kappa_p=2.79$ for the proton target, and
mass parameter $\Lambda_p$ to fit to data.

The $\omega NN$ coupling constants, $g^v_{\omega NN}=15.6$ and
$g^t_{\omega NN}=0$, are taken for consistency with other hadron
reactions.
The nucleon isoscalar form factor at the $\omega NN$ vertex in the 
$u$-channel is now
constructed in terms of quark densities for the proton and neutron charge
form factors \cite{guidal},
\begin{eqnarray}
\label{eqn:isoff}
\hspace{6em}F^{(s)}(u)={\frac{1}{3}}\left[u(u)+d(u)\right],
\end{eqnarray}
where the unpolarized $u$ and $d$ quark
distributions are given as the first moments of their respective GPDs,
\begin{eqnarray}\label{eqn:quark}
\hspace{4em}
q(u)=\int_0^1 dx \,H^q(x) x^{-(1-x)\alpha'_qu}
\end{eqnarray}
with the slope $\alpha'=0.9$ GeV$^{-2}$ chosen and the GPDs for the valence quarks $u_v$ and $d_v$ taken from Ref.~\cite{martin2002}. In what follows, the GPDs are understood as zero-skewness.  

\begin{figure}[ht]
\centering
\includegraphics[width=0.46\textwidth]{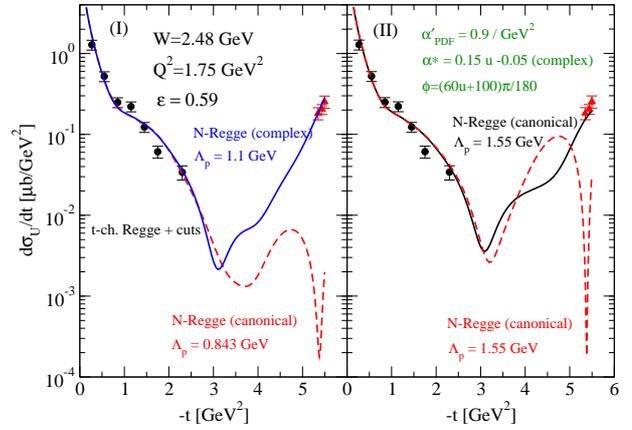}
\caption{Unpolarized cross section, $d\sigma_U/dt$, for $\omega$ electroproduction from the
conventional Regge model without parton contribution (left panel) 
and from the extended model with parton
contributions (right panel). Data are the same as
in Fig.~\ref{fig:omega_t_slope}. In (I) the nucleon exchange with 
the on-shell cutoff $\Lambda_p=0.843$~GeV falls far below the data points.
In (II) the
nucleon exchange in (I) is raised up with the large cutoff 
$\Lambda_p=1.55$~GeV, but cannot reach over the data yet.
Details about figure captions are given in the text. Data are
taken from Refs. \cite{Morand05,Li:2019xyp}} \label{fig:regge1}
\end{figure}

The unpolarized cross section, $d\sigma_U/dt$, is presented in Fig. \ref{fig:regge1} at energy $W=2.48$~GeV with photon virtuality
$Q^2=1.75$~GeV$^2$, and polarization $\varepsilon=0.59$. The
CLAS-6~GeV data below $-t<3$ GeV$^2$ \cite{Morand05} are
reproduced by the $t$-channel Regge model with cuts in Ref.
\cite{yu2} extended to electroproduction. The cross section of the red
dashed curve in the panel (I) shows the conventional description of the nucleon Reggeon which chooses the canonical phase with
the cutoff $\Lambda_p=0.843$ GeV for the on-shell $\gamma^*NN$ 
form factors in Eqn. (\ref{eqn:dipoleff}), but excludes the parton contribution in Eqn. (\ref{eqn:nucleon2}). 
To agree with data at $-t\approx5.5$
GeV$^2$, we raise up the cross section by using the complex phase
and the large cutoff mass $\Lambda_p=1.1$ GeV to obtain the blue
solid curve. This means that the current data set can be
described in the conventional approach without partonic
contributions.

Panel (II) of Fig. \ref{fig:regge1} shows the difference between
the solid curve with the parton contribution and the dashed one
without it in Eqn. (\ref{eqn:nucleon2}), given the canonical phase
for the nucleon trajectory and cutoff $\Lambda_p=1.55$ GeV.  
For the modified nucleon Reggeon $\widetilde{\cal R}^N$ in Eqn.
(\ref{eqn:nucleon2}), the trajectory $\widetilde\alpha_N(u)=0.15u-0.05$
and the complex phase are taken with the relative phase adjusted to
$\phi(u)=(60u+100)\pi/180^\circ$.

\begin{figure}
\includegraphics[width=0.48\textwidth]{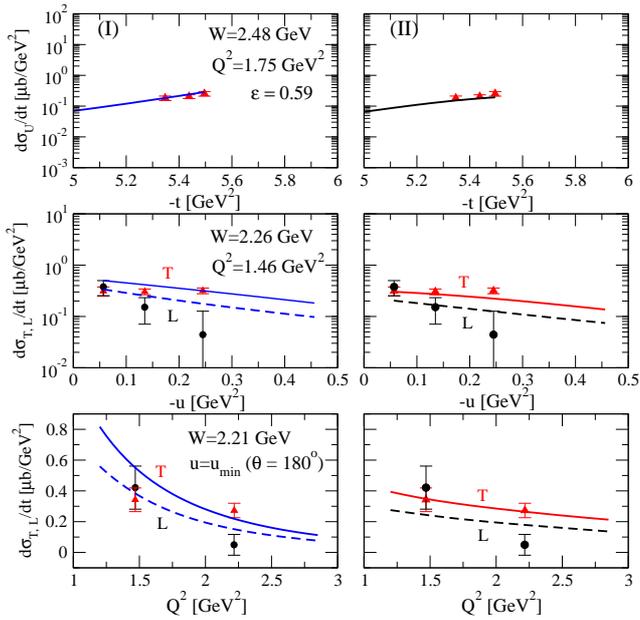}
\caption{ The unpolarized, transverse, and longitudinal cross
sections for $\omega$ electroproduction from the conventional Regge approach (left column) and from the hybrid model
with parton contributions added (right column). The parameters for both
models are the same as in Fig. \ref{fig:regge1}. }
\label{fig:regge2}
\end{figure}

Figure \ref{fig:regge2} presents the same cross section $d\sigma_U/dt$ with Fig. \ref{fig:regge1} magnified in the interval $5<-t<6$ GeV$^2$. The $u$ and $Q^2$-dependence of transverse and longitudinal cross sections $d\sigma_T$ and
$d\sigma_L$ are shown in the middle and lower panels, respectively. With the same notation for panels (I) and (II)
as in Fig. \ref{fig:regge1},
both models yield plausible descriptions of the $F_\pi$
Collaboration data, respectively, though each $d\sigma_T$ and $d\sigma_L$
still lack accuracy. Therefore, for
clarity of theory, more data are needed in the
range of $-t$ below 5.3 GeV$^2$.

\begin{figure}[ht]
\centering
\includegraphics[width=0.9\hsize]{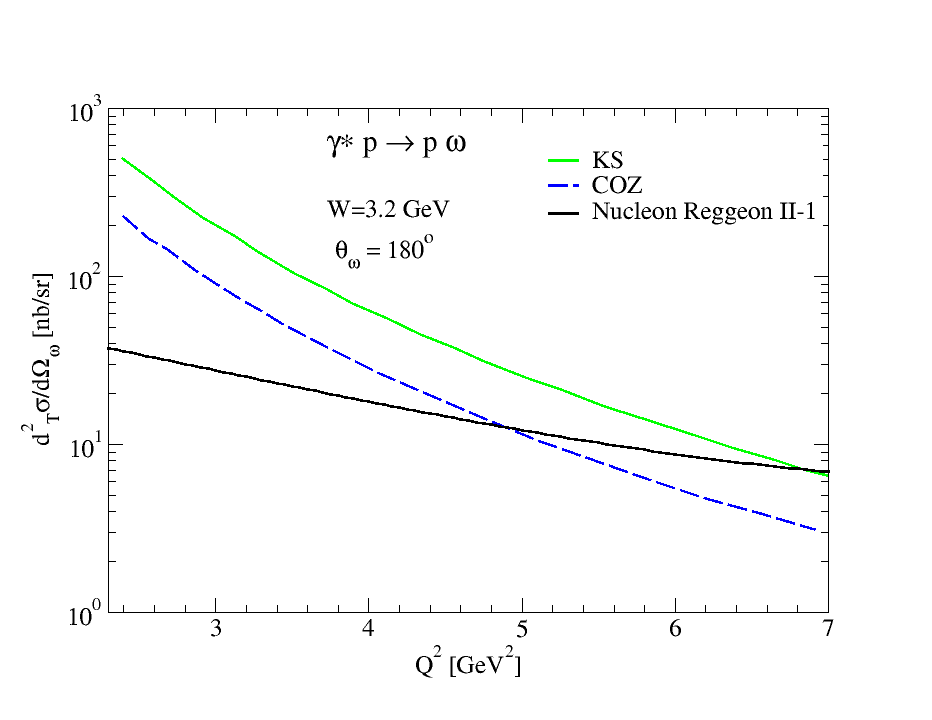}%
\caption{ $Q^2$-dependence of the cross section $d^2\sigma_T/d\Omega$ for $\omega$ electroproduction
from the collinear QCD description with TDAs, the conventional Regge model and the Reggeon with parton model. 
The conventional approach without parton contributions  deviates much 
from others at high $Q^2$.}
\label{fig:regge3}
\end{figure}

In the sense that the data are reproduced in the Regge model without parton 
contributions, it is interesting to compare the conventional Regge model in Fig.~\ref{fig:regge1} (I) with the result from JML18 model presented in Fig.~\ref{fig:omega_t_slope} in Sec.~\ref{sec:hallc_omega},
which comes from the $u$-channel cuts to reproduce the cross section with the standard form factor and canonical phase preserved for the nucleon.

It is worth mentioning the comparison between hadron models and
QCD inspired calculation for the electroproduction cross section.
According to the TDA calculation, the $Q^2$-dependence of
the cross-section shows  $1/Q^8$ scaling. Fig. \ref{fig:regge3}
compares this TDA prediction for the cross-section to the nucleon Reggeon exchange model
with and without parton contributions.
At higher $Q^2$, the Regge plus parton model agrees with TDA models, but the 
conventional model shows a steeper  $Q^2$-dependence.   
At low $Q^2$, there is a significant inconsistency between Regge models and 
TDA predictions which should be clarified by experimental measurements.

\begin{figure}[ht]
\centering
\includegraphics[width=0.44\textwidth]{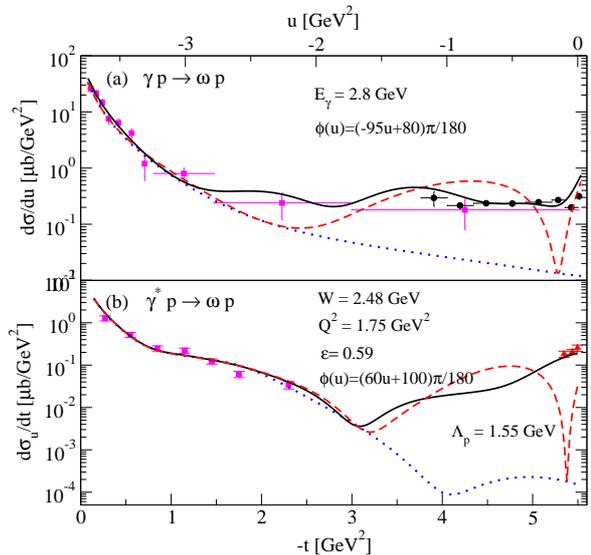}%
\caption{Cross sections for $\omega$ photoproduction  (a) and electroproduction
(b) at similar energies $W=2.48$ GeV. In both panels the
dotted curve depicts the meson Regge-pole exchange in the $t$-channel.
The dashed curve in both panels represents the nucleon Reggeon without partons as in Fig. \ref{fig:regge1}.} \label{fig:regge4}
\end{figure}

A few closing remarks are in order to close the discussed ``Regge'' approach. According to the
extrapolation of $\omega$ electroproduction data at 6 GeV  to real
photon point $Q^2=0$, it is found that the cross section
$\sigma_T$ at the real photon limit is consistent with the Regge
model prediction \cite{yu} at $E_\gamma=2.48$ GeV and $u\simeq
u_{min}$ in the backward direction. This is well reproduced in Fig. \ref{fig:regge4}.  Nevertheless, the data are
insufficient to clarify whether or not a nucleon dip exists in the
case of electroproduction. Thus, it may be one of the important
tasks to collect more data interpolating to the mid angle region
in future experiments. 
Within the Regge framework in Eqn. (\ref{eqn:nucleon2}), 
the nucleon exchange in photoproduction as well as in electroproduction 
shows the expected dip at the nonsense zeros with
the canonical phase applied to the nucleon trajectory. Moreover, as
the cross section at backward angles is believed to be the sum of
the nucleon and parton contributions, the cross section for
electroproduction exhibits the roles of partons around the dip region
quite similar to the case of photoproduction. However, 
this coincidence is not necessarily true. One could obtain the different 
electroproduction cross section with the different relative phase
$\phi(u)$ between nucleon and parton which is nevertheless able to
reproduce the $F_\pi$ data. 
Therefore, more data such as the beam polarization 
could serve to discriminate the interference pattern between the nucleon 
and partons within the current approach.

\subsection{Data Driven Predictions of Backward Two-Meson Photoproduction}
\label{sec:two_meoson_photoproduction}

The upcoming experiments at Jefferson Lab Hall C (E12-20-007 \cite{E12-20-007}, E12-19-006 \cite{E12-19-006} and E12-09-011 \cite{E12-09-011}) will measure the differential cross section for exclusive $\pi^0$, $\omega$ and $\phi$ meson production in the backward direction. For the case of the meson resonances like the $\rho$, $\omega$ or $\phi$ one anticipates subsequent decays to pions or kaons, since the branching fraction is generally largest for these processes. Thus $\gamma p\to \pi\pi p$, $\gamma p\to \pi\pi\pi  p$ and $\gamma p\to K K p$ are important contributions to the non-resonant background of the above $2\to2$ processes. Here we describe an approach which utilizes information from related two-body final states to constrain the dynamics of the processes of interest, leading to an effectively parameter free estimation of the non-resonant background. Such an approach has been used in the description of forward $\gamma p\to \pi\pi p$. Here we discuss how it may be applied to backward scattering.

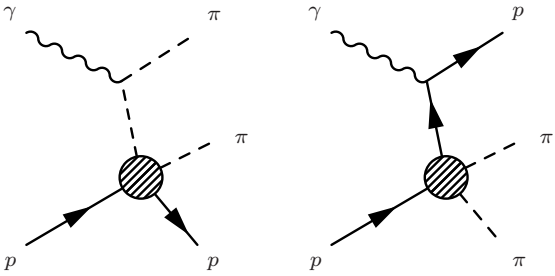
\begin{figure}
\vspace{10pt}
\begin{fmffile}{feynDiags3}
\begin{fmfgraph*}(80,80)
\fmfleft{l1,l2}
\fmfright{r1,r2,r3}
\fmf{fermion}{l1,v1,r1}
\fmf{photon}{l2,v2}
\fmf{dashes}{v1,r2}
\fmf{dashes}{v1,v2,r3}
\fmfblob{0.2w}{v1}
\fmflabel{$p$}{l1}
\fmflabel{$\gamma$}{l2}
\fmflabel{$p$}{r1}
\fmflabel{$\pi$}{r2}
\fmflabel{$\pi$}{r3}
\end{fmfgraph*}
\end{fmffile}
\begin{fmffile}{udeck}
\hspace{20pt}
\begin{fmfgraph*}(80,80)
\fmfleft{l1,l2}
\fmfright{r1,r2,r3}
\fmf{fermion}{l1,v1}
\fmf{dashes}{v1,r1}
\fmf{photon}{l2,v2}
\fmf{dashes}{v1,r2}
\fmf{fermion}{v1,v2,r3}
\fmfblob{0.2w}{v1}
\fmflabel{$\pi$}{r2}
\fmflabel{$\pi$}{r1}
\fmflabel{$p$}{r3}
\fmflabel{$p$}{l1}
\fmflabel{$\gamma$}{l2}
\end{fmfgraph*}
\end{fmffile}
\vspace{10pt}
\vspace{10pt}
\caption{The dominant Deck Mechanism for two pion photoproduction for small $t_i$ (left) and  the analogous process for small $u$ (right). }\label{fig:deckMechanism}
\end{figure}

The procedure to describe backward production of mesons follows closely their forward production, which we first describe. 
In the kinematic limit where $|t_\pm|=|(q-k_{\pi_\pm})^2|$ is small, the largest contribution to the non-resonant background may be described using the so-called \textit{Deck mechanism}~\cite{Deck:1964hm,Pumplin:1970kp}. While this has seen good success in forward kinematics~\cite{Bibrzycki:2018pgu}, the application of this approach to backward kinematics has been less studied. The essential idea which underlies the Deck mechanism is that at low $|t_\pm|$, the incoming photon scatters diffractively from the target nucleon. As a result of the long-range character of the interaction, it is expected that pion exchange will dominate in this kinematic region. The essential idea which underlies the Deck mechanism is that at low $|t_\pm|$, the incoming photon dissociates into a $\pi^+\pi^-$ pair with one of pions scattering diffractively from the target nucleon. As a result of the long-range character of the interaction, it is expected that pion exchange will dominate in this kinematic region. Thus one expects the process to proceed via one-pion exchange with the pion's virtuality given by $t_\pm$. This is shown explicitly in Fig.~\ref{fig:deckMechanism}. Symbolically, the Deck mechanism for this process is
\begin{equation}
\hspace{1em}\mathcal{M}=\frac{\mathcal{M}(\gamma\to \pi+\pi^*)\times\mathcal{M}(p+\pi^*\to p+\pi)}{t_i-m_\pi^2}
\end{equation}
where $\mathcal{M}(\gamma\to \pi+\pi^*)$ denotes the amplitude for the dissociation of the incident photon into a pair of pions, and $\mathcal{M}(p+\pi^*\to p+\pi)$ denotes the $p+\pi^*\to p+\pi$ scattering amplitude. The strength of such a description lies in the possibility of using information about the on-shell $p+\pi\to p+\pi$ scattering amplitudes, for which there now exists a wealth of experimental data, and theoretically motivated parameterizations~\cite{Workman:2012hx,Mathieu:2015gxa}. It is natural to consider a generalization of the Deck mechanism to $u$-channel backward scattering. In this kinematic region, the analogous process arises from the exchange of low-mass baryons in the $u$-channel. Symbolically, the corresponding amplitude is
\begin{equation}
\hspace{2em}\mathcal{M}=\frac{\mathcal{M}(\gamma\to p+\overline{p}^*)\times\mathcal{M}(p+\overline{p}^*\to \pi+\pi)}{u-m_N^2}
\end{equation}
In this case, the required empirical input are the scattering amplitudes for $N\overline{N}\to\pi\pi$. One can also generalize this approach to produce a description of two-kaon photoproduction, where the required two-body amplitudes are $N\overline{N}\to KK$. Such a three-body final state is relevant for the study of backward $\phi$ photoproduction. 

\subsection{From GPDs to TDAs}
\label{sec:TDA}

Generalized parton distributions (GPDs) provide a modern description of the complex internal structure of the nucleon, which gives access to the correlations between the transverse position and the longitudinal momentum distribution of quarks and gluons in the nucleon. 

\begin{figure} 
\centering
  \includegraphics[width=0.35\textwidth]{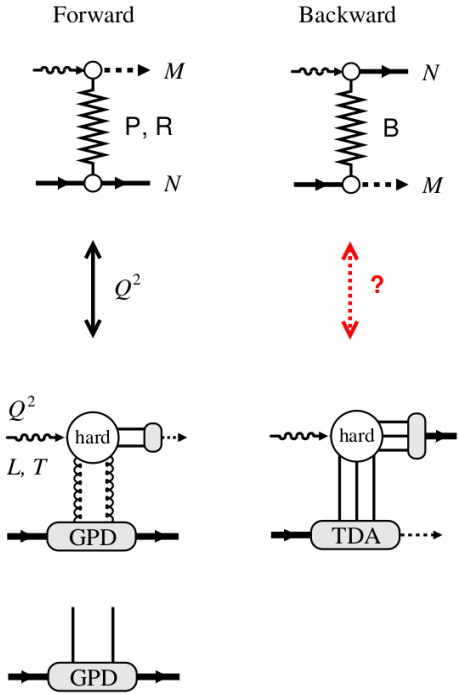}
\caption{Forward and backward collinear factorization schemes. }
\label{fig:transition_2} 
\end{figure}

In addition, GPDs give access to the orbital momentum contribution of partons to the spin of the nucleon~\cite{Ji:1997}. 
The possibility of the direct experimental 
access to GPDs still remains controversial, see discussion e.g. in Refs. 
\cite{Ji:2004,Kumericki:2009uq,Bertone:2021yyz}.
The prime experimental channels for studying the GPDs are through the DVCS (Deeply Virtual Compton Scattering) and DEMP (Deep Exclusive Meson Production)
processes. An example DEMP reaction, 
\begin{equation}
\hspace{3em}\gamma^* (q) + p (p) \rightarrow p (p+\Delta) + M(q-\Delta),
\end{equation}
is shown in the left-lower section of Fig.~\ref{fig:transition_2}. Here, a leading meson $M$ is produced in forward-angle kinematics, and $\Delta$ is the $t$-channel momentum transfer, recall $\Delta^2 = t$.

 In order to access the GPD colinear factorization regime, the kinematic variable requirements are as follows: sufficiently high $Q^2$, large $s$, fixed $x_B$ and $t\rightarrow0$~\cite{Ji:2004}. Here, the definition of ``sufficiently high $Q^2$'' is a process-dependent terminology. 
Based on the DVCS experimental data~\cite{FX:2008, Camacho:2006, Defurne:2015},  it has been shown that the range of ``sufficiently high $Q^2$'' for factorization to apply to this reaction lies between 1 and 5 GeV$^2$; this is sometimes referred to as ``early scaling''~\cite{Voutier:2009}. Note that, the most recent data from JLab 12 GeV ~\cite{Bhetuwal:2021} has covered the upper range of the early scaling region.

Under the Colinear Factorization (CF) regime, a parton is emitted from the nucleon, interacts with the incoming virtual-photon, then returns to the nucleon after the interaction~\cite{Ji:2004}. Studies~\cite{Frankfurt:2011cs, Kroll:2016} have shown that perturbative calculation methods can be used to calculate the scattering process and extract GPDs from data, while preserving the universal description of the hadronic structure in terms of QCD principles. One limitation is that GPD factorization requires $t \sim t_{min}$, namely, the process defaults to a fast-meson and slow-nucleon final state. Processes such as the one discussed in this $u-$channel workshop cannot be correctly accounted for by such a description.

 The proof of CF for meson exclusive production~\cite{Collins:1997} is essentially based on the observation that the cancellation of the soft gluon interactions is intimately related to the fact that the meson arises from a quark-antiquark pair generated by the hard scattering. Thus, the pair starts as a small-size configuration and only substantially later grows to a normal hadronic size, and thus a meson. Similarly, it has been advocated~\cite{Frankfurt:1999fp} that the factorization theorem should also be valid for the production of leading baryons
\begin{equation}
\hspace{3em}\gamma^{*} (q) + p (p) \rightarrow B(q-\Delta) + M (p+\Delta) \label{eqn:leading_baryon} \end{equation}
and even leading antibaryons ($\bar{B}$)
\begin{equation}
\hspace{3em}\gamma^{*} (q) + p (p) \rightarrow \bar{B}(q-\Delta) +  B_2 (p+\Delta) \label{eqn:bbar}
\end{equation}
where $B_2$ is a system with baryon number of two. Here, $\Delta$ is the $u$-channel momentum transfer, recall $\Delta^2 = u$.

To describe reactions~(\ref{eqn:leading_baryon})
within the CF framework it turns out
necessary to introduce
%
$B \to M$ transition  matrix elements
of three-quark light cone operators, the TDAs 
(which were called super-SPDs in~\cite{Frankfurt:2002}), 
\begin{eqnarray}
\nonumber
&&\hspace{5em} \int\prod_{i=1}^{3}dz^{-}_{i}\exp[i\sum_{i=1}^{3}x_i (p\cdot z_i)] \cdot\\\nonumber
&&\langle M(p+\Delta)|\\\nonumber
&&\hspace{2em}|\varepsilon_{c_1 c_2 c_3}\Psi^{c_1}_{j_1}(z_1)\Psi^{c_2}_{j_2}(z_2)\Psi^{c_3}_{j_3}(z_3)|N(p)\rangle\biggr\vert_{z^{+}_i = z^{\perp}_i=0}\\
&& = \delta(2\xi -x_1 - x_2 - x_3) F_{j_1j_2j_3} (x_1, x_2, x_3, \xi, u),
\end{eqnarray}
where 
$c_{1,2,3}$
are the color indices, $j_{1,2,3}$ are spin-flavor indices, and $F_{j_1 j_2 j_3} (x_1, x_2, x_3 , \xi, u)$ are the new objects which can be decomposed into invariant spin-flavor structures depending on the quantum numbers of the meson $M$. They turn to be functions of the momentum fraction variables $x_i$%
the skewness parameter $\xi =  \Delta^+/(2p^++\Delta^+)$ which characterizes the longitudinal momentum transfer between the initial state nucleon and the final state meson, and momentum transfer squared $u = \Delta^2$. 

To ensure an early onset of scaling it is natural to consider the process as a function of $Q^2$ at fixed $\xi$ and $u$. If  color transparency~\cite{Farrar:1988me,Jennings:1990ma,Jain:1995dd} suppresses the final state interaction between the fast moving nucleon and the residual meson state early enough, then a qualitative prediction~\cite{frankfurt2000} from this description is that the cross section ratio between the reaction~(\ref{eqn:leading_baryon}) and the elastic $eN$ scattering one can be written as: 
$$
\frac{\sigma(eN \rightarrow eNM)}{\sigma(eN \rightarrow eN)} \sim f_M(\alpha_M, p_t)(1 - \alpha_M)\,
$$
where 
$$\alpha_M = \frac{p_{M^-}}{p_{N^-}} = \frac{1-\xi}{1+\xi} \,$$
and $(1-\alpha_M)$ is a flux factor.

In the case of pion production, the soft pion limit, corresponding to $\alpha_M \sim m_\pi / m_N$, $p_t \le m_\pi$, is of special interest because one can use the factorization theorem and chiral perturbation theory in a manner similar to the considered process $eN\rightarrow eN\pi$ at large $Q^2$ and small $W$~\cite{Pobylitsa:2001}. However, reaching this kinematic regime may require extremely high $Q^2$. 

\subsection{Collinear QCD factorization and  baryon-to-meson TDAs}
Five years after the remarks~\cite{Frankfurt:1999fp} extending the GPD factorization concepts to reactions with baryonic exchange, the relevance of baryon-to-meson matrix elements of 
three-quark operators on the light cone was rediscovered~\cite{Pire:2004ie,Pire:2005ax} in an effort to draw a consistent partonic picture of various electromagnetic processes accessible in antiproton-nucleon annihilation within the $\overline{\textrm{P}}$ANDA program~\cite{Lutz:2009ff} and of backward electroproduction processes.

The baryon-to-meson TDAs (see \cite{Pire:2021hbl} for a review) are defined  within the QCD collinear factorisation scheme through matrix
elements of the non-local three-quark (antiquark) operators on the light cone 
($n^2=0$):
\begin{eqnarray}
\hspace{2em}&&\hat{O}^{\alpha \beta \gamma}_{\rho \tau \chi}( \lambda_i n)
 =\\\nonumber
&&\hspace{2em}\varepsilon_{c_{1} c_{2} c_{3}}
\Psi^{c_1 \alpha}_\rho(\lambda_1 n)
\Psi^{c_2 \beta}_\tau(\lambda_2 n)
\Psi^{c_3 \gamma}_\chi (\lambda_3 n)
\label{eqn:operators}
\end{eqnarray}
between an initial baryon state with momentum $p_N$ and a final meson state with momentum $p_M$, where
$\alpha$, $\beta$, $\gamma$
stand for the quark (antiquark) flavor indices and
$\rho$, $\tau$, $\chi$
denote the Dirac spinor indices. Antisymmetrization is performed over the
color group indices
$c_{1,2,3}$. 
Gauge links in
(\ref{eqn:operators}) are omitted by adopting the light-like axial gauge
$A \cdot n=0$.
These non-perturbative objects
share common features both with baryon distribution amplitudes (DAs), introduced in \cite{Lepage:1980,Chernyak:1983ej} as baryon-to-vacuum matrix elements of the same operators
(\ref{eqn:operators}),
and with generalized parton distributions (GPDs), since the matrix element in question depends on the longitudinal momentum transfer
$\Delta^+=(p_{ M}-p_N) \cdot n$
between a baryon and a meson characterized by the skewness variable
$
\xi= -\frac{(p_{M}-p_N) \cdot n}{(p_{ M}+p_N) \cdot n}
$
and by a transverse momentum transfer
$\vec \Delta_T$.

The QCD evolution equations obeyed by baryon-to-meson TDAs  distinguish the
Efremov-Radyushkin-Brodsky-Lepage (ERBL)-like domain, in which all three momentum
fractions of quarks
are positive, and two  Dokshitzer-Gribov-Lipatov-Altarelli-Parisi (DGLAP) -like regions,
in which either one or two momentum fractions of quarks are negative.

\begin{figure*} 
\centering
  \includegraphics[width=0.95\textwidth]{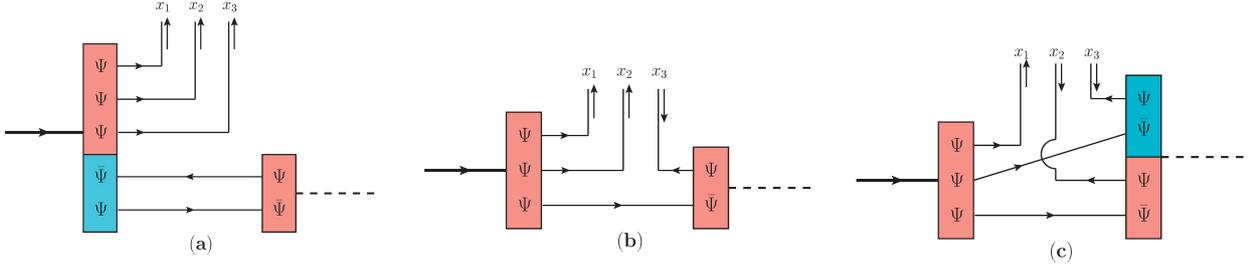}
\caption{Interpretation of nucleon-to-meson TDAs at low normalization scale: {\bf  (a):} Contribution in the ERBL region (all $x_i$ are positive);
    {\bf  (b):} Contribution in the DGLAP I region (one  $x_i$  is negative).
    {\bf  (c):} Contribution in the DGLAP II region (two  $x_i$  are negative).
     }
\label{fig:TDA} 
\end{figure*}

For simplicity, let us from now on focus on the nucleon-to-meson case. The physical picture encoded in nucleon-to-meson TDAs is conceptually close to that
contained in nucleon GPDs and nucleon DAs. Nucleon-to-meson TDAs characterize partonic correlations inside a nucleon and give access to the momentum distribution
of the baryonic number inside the nucleon. The same operator also defines the nucleon DA,
which can be seen as a limiting case of nucleon-to-meson TDAs with the meson state replaced by the vacuum.
In the language of the Fock state decomposition, nucleon-to-meson TDAs are not restricted to the lowest Fock state
as DAs. They rather probe the non-minimal Fock components with additional
quark-antiquark pairs:
\begin{eqnarray}
\hspace{2em}&&
| {\rm Nucleon} \rangle= |\Psi \Psi \Psi \rangle+ |\Psi \Psi \Psi; \,  \bar{\Psi} \Psi \rangle+....\; \nonumber\\
&&
| {\rm Meson} \rangle= |\bar{\Psi}\Psi \rangle+ |\bar{\Psi}\Psi; \, \bar{\Psi} \Psi \rangle+....\;
\end{eqnarray}
depending on the particular support region in question
(see Fig.~\ref{fig:TDA}). Note that this interpretation can be
justified only at a very low normalization scale and can be
significantly altered at higher scales due to the evolution effects.

Similarly to GPDs, by Fourier transforming nucleon-to-meson TDAs to the
impact parameter space
($\vec \Delta_T \to \vec b_T$),
one obtains additional insight into the nucleon structure in the transverse plane \cite{Pire:2019nwa}.
This allows one to perform the femto-photography of hadrons
\cite{Ralston:2001xs,Pire:2019nwa}
from a new perspective.

For a given flavor contents ({\em e.g.} 
$uud$ proton-to-$\pi^0$ TDA), the parametrization
of the leading twist-$3$
$\pi N$
TDA involves
eight
invariant functions,
each depending on the
three
longitudinal momentum fractions
$x_i$ that are subject to the momentum conservation constraint $\sum_{i=1}^3 x_i=2 \xi$,
where $\xi$ is the skewness parameter,
momentum transfer squared
$\Delta^2$
as well as on the factorization scale
$\mu^2$:
\begin{eqnarray}
&&
4 (p \cdot n)^3 \int \left[ \prod_{j=1}^3 \frac{d \lambda_j}{2 \pi}\right]
e^{i \sum x_k \lambda_k (p \cdot n)}
 \langle     \pi^0(p_\pi)|\,  \varepsilon_{c_1 c_2 c_3} \nonumber\\&&
 u^{c_1}_{\rho}(\lambda_1 n)
u^{c_2}_{\tau}(\lambda_2 n)d^{c_3}_{\chi}(\lambda_3 n)
\,|N^p(p_N,s_N) \rangle 
\nonumber \\ &&
= \delta(\sum x_i-2 \xi) i \frac{f_N}{f_\pi}\Big[  V^{(p\pi^0)}_{1}(x_{i}, \xi,\Delta^2)  (  \hat{p} C)_{\rho \tau}(U^+)_{\chi}
\nonumber \\ &&
+A^{(p\pi^0)}_{1}(x_{i}, \xi,\Delta^2)  (  \hat{p} \gamma^5 C)_{\rho \tau}(\gamma^5 U^+ )_{\chi}
\nonumber \\ &&
 +T^{(p\pi^0)}_{1}(x_{i}, \xi,\Delta^2)  (\sigma_{p\mu} C)_{\rho \tau }(\gamma^\mu U^+ )_{\chi}
 \nonumber \\ &&
 + m_N^{-1} V^{(p\pi^0)}_{2} (x_{i}, \xi,\Delta^2)
 ( \hat{ p}  C)_{\rho \tau}( \hat{\Delta}_T U^+)_{\chi}\nonumber \\&&
+ m_N^{-1}
 A^{(p\pi^0)}_{2}(x_{i}, \xi,\Delta^2)  ( \hat{ p}  \gamma^5 C)_{\rho\tau}(\gamma^5  \hat{\Delta}_T  U^+)_{\chi}
  \nonumber \\ &&
+ m_N^{-1} T^{(p\pi^0)}_{2} (x_{i}, \xi,\Delta^2) ( \sigma_{p\Delta_T} C)_{\rho \tau} (U^+)_{\chi}
\nonumber \\&&
+  m_N ^{-1}T^{(p\pi^0)}_{3} (x_{i}, \xi,\Delta^2) ( \sigma_{p\mu} C)_{\rho \tau} (\sigma^{\mu\Delta_T}
 U^+)_{\chi}
 \nonumber \\ &&
+ m_N^{-2} T^{(p\pi^0)}_{4} (x_{i}, \xi,\Delta^2)  (\sigma_{p \Delta_T} C)_{\rho \tau}
(\hat{ \Delta}_T U^+)_{\chi} \Big]
 \label{eqn:Old_param_TDAs}
\end{eqnarray}
Here,
$f_\pi=93$ MeV
is the pion weak decay constant, $m_N$ the nucleon mass and
$f_N$
determines the value of the nucleon wave function at the origin;
$\hat{v} \equiv v_\mu \gamma^\mu$;
$\sigma^{\mu\nu}= \frac{1}{2} [\gamma^\mu, \gamma^\nu]$; $\sigma^{v \mu} \equiv v_\lambda \sigma^{\lambda \mu}$;
$C$
is the charge conjugation matrix and
$U^+= \hat{p} \hat{n} \, U(p_N,s_N)$
is the large component of the nucleon spinor. 

A simple model for 
$\pi N$ 
TDAs suggested in Ref.~\cite{Pire:2011xv}
accounts for the contribution of 
the cross-channel
nucleon exchange. This model is conceptually similar to the commonly used
pion exchange model for the polarized nucleon GPD
$\tilde{E}$.
With the use of the
$\pi N$
TDA parametrization
(\ref{eqn:Old_param_TDAs})
the nucleon pole model reads:
\begin{eqnarray}
&&
\big\{ V_1, \, A_1 , \, T_1  \big\}^{p \pi^0} ( 
x_i, \xi,\Delta^2,\, \mu^2)\Big|_{N(940)}
\nonumber \\ &&
 =\frac{\Theta_{\rm ERBL}(x_k)}{2 \xi } \times 
  \frac{g_{\pi NN} \,m_N f_\pi}{\Delta^2-m_N^2}   \frac{1-\xi}{1+\xi}
 \big\{ V^p,\,A^p, \,T^p  \big\}\left( \frac{x_i}{2 \xi}
 \right);
 \nonumber \\  &&
 \big\{ V_2, \, A_2 , \, T_2, \, T_3  \big\}^{p \pi^0} ( 
x_i, \xi,\Delta^2, \, \mu^2)\Big|_{N(940)}
\nonumber \\ &&
=\frac{\Theta_{\rm ERBL}(x_k)}{2 \xi } \times    \frac{g_{\pi NN} \, m_N f_\pi}{\Delta^2-m_N^2}    
 \big\{ V^p,\,A^p, \,T^p, \, T^p  \big\}\left( \frac{x_i}{2 \xi}
 \right);
\nonumber \\  &&
   T_4^{p \pi^0 } ( 
x_i, \xi,\Delta^2, \, \mu^2)\Big|_{N(940)}=0;
\nonumber \\  &&
\big\{ V_{1,2}, \, A_{1,2} , \, T_{1,2,3,4}  \big\}^{p \pi^+} ( 
x_i, \xi,\Delta^2, \, \mu^2)\Big|_{N(940)}\nonumber\\&&
= -\sqrt{2} \big\{V_{1,2}, \, A_{1,2} , \, T_{1,2,3,4}    \big\}^{p \pi^0 } ( 
x_i, \xi,\Delta^2,\, \mu^2)\Big|_{N(940)}\,
\label{eqn:Nucleon_exchange_contr_VAT}
\end{eqnarray}
Here
$
\Theta_{\rm ERBL}(x_k)  \equiv  \prod_{k=1}^3 \theta(0 \le x_k \le 2 \xi)
$
ensures the pure ERBL-like support of TDAs and
$g_{\pi NN}\approx 13$
is the pion-nucleon phenomenological coupling.
This turns out to be a consistent model for
$\pi N$
TDAs in the ERBL-like region and satisfies the polynomiality conditions 
and the appropriate symmetry relations.

By an obvious change of couplings, the model
(\ref{eqn:Nucleon_exchange_contr_VAT})
can be generalized to the
case of other light mesons
($\eta$, $\eta'$, $K$, ... {\it etc.}).
Also it is not necessarily limited to the contribution
of the cross-channel nucleon exchange.
In Ref.~\cite{Pire:2011xv}
the contribution of the cross-channel
$\Delta(1232)$ exchange
into
$\pi N$
TDAs was worked out explicitly. Finally, 
the cross-channel nucleon exchange model was generalized for the case of
nucleon-to-vector meson TDAs \cite{Pire:2015kxa}.

The aforementioned baryon-to-meson TDA model describes TDAs
only within the ERBL-like support region. To get a model defined
on the complete support domain one may rely on the spectral representation
for baryon-to-meson TDAs in terms of quadruple distributions suggested in Ref.~\cite{Pire:2010if}:
\begin{eqnarray}
\label{Spectral_for_GPDs_x123}
&&
H^{({ M} N)}(x_1,\,x_2,\,x_3=2 \xi -x_1-x_2,\,\xi,\,t) = \\ &&
\left[
\prod_{i=1}^3
\int_{\Omega_i} d \beta_i d \alpha_i
\right] f(\beta_i, \, \alpha_i,\,t) \delta(\sum_i \beta_i) \nonumber\\&&
\delta(x_1-\xi-\beta_1-\alpha_1 \xi) \,
\delta(x_2-\xi-\beta_2-\alpha_2 \xi) 
\delta(\sum_i \alpha_i+1)
 \,, 
\nonumber
\end{eqnarray}
where
$\Omega_{i}= \{|\beta_i| \le 1; \, |\alpha_i| \le 1- |\beta_i|\}$
denote three copies of the usual domain in the spectral parameter space.
The spectral density
$f$
is an arbitrary function of six variables, which are  subject to two constraints, 
and therefore effectively is a quadruple distribution.
Similarly to the familiar double distribution representation for GPDs  a
quadruple distribution representation
for TDAs turns to be the most
general way to implement the support properties of TDAs as well as the
polynomiality property for the $x_i$-Mellin moments,
which is a direct consequence of Lorentz invariance
(see Ref.~\cite{Lansberg:2011aa}).

Contrarily to GPDs, TDAs do not possess the comprehensive forward limit
$\xi \to 0$.
This complicates the construction of  phenomenological ans\"atze for quadruple
distributions.
To calculate a scattering amplitude in this framework, the TDAs are to be convoluted with a hard amplitude $A_H$ calculated in perturbative QCD. this hard  amplitude  has the same scaling behaviour in $Q^2$ as the nucleon form factor ($A_H \sim Q^{-4}$ ), leading for instance to $\sigma_T \propto Q^{-8}$ in electroproduction cross sections (see section 7.1 of \cite{Pire:2021hbl}).

Without entering into too much detail on the phenomenology of $u$ channel reactions within the TDA framework, the recent experimental data from JLab \cite{Park:2017irz,Li:2019xyp,Diehl:2020uja}   gives hope that leading twist dominance for backward meson electroproduction may already be witnessed at moderate values (a few GeV$^2$) of the hard scale $Q^2$. The estimates of cross sections presented  in  \cite{Lansberg:2012ha,Pire:2013jva} for the exclusive  $\pi^0$  production produced through the $p \bar p$ annihilation processes, $\bar p \;p \to \gamma^* \pi^0$ and
$\bar p \;p \to J/\Psi \pi^0$ indicate that they can be measured at $\overline{\textrm{P}}$ANDA/FAIR. Let us stress the well-known fact that the onset of the scaling region at some value of the large momentum transfer $Q$ is a process-dependent feature that should be studied for every reaction.

In summary, the TDA framework is a way to study a new class
of hard exclusive reactions that can be used as a testbed for the available theoretical tools that have been so successful to describe and understand hard inclusive, semi-inclusive and forward exclusive physics. Since we are at the beginning of this new physics domain, we have to deal with extending the proofs of factorization theorems and with estimating possible non-factorized contributions to the scattering amplitudes. We shall also have to consider QCD evolution effects in order to master higher twist and next-to-leading order in $\alpha_s$ corrections in a consistent way. Moreover, we shall have to design suitable physical observables for which one may expect early onset of scaling behavior. Since we believe that our understanding 
of QCD has reached a high level of maturity, we can and must challenge all these issues to refine the existing and develop
new theoretical methods.      
With the help of precise and various experimental data, we will thus expect to gain new specific physical information on the baryonic structure. 
Here we can mention several points:
\begin{itemize}
\item Baryon-to-meson (and baryon-to-photon) TDAs, which may be extracted from backward meson electroproduction, contain more information
on the hadronic structure than baryon distribution amplitudes. In particular,
we may gain access to the non-minimal components of hadronic light-cone wave functions
($5$-quark component for baryons and $q \bar{q} q \bar{q}$ for mesons). 
Addressing different mesons ($\pi$, $\eta$, $\eta'$, $\rho$, $\omega$, $\phi$, $K$),
we can probe components with different flavor and helicity content.

\item The impact parameter picture developed for baryon-to-meson (and baryon-to-photon) TDAs
is conceptually close to that for GPDs. In particular, it  provides access to the baryon charge distribution in the transverse plane. It can also provide means to study the meson cloud (and electromagnetic cloud) inside baryons (particularly the nucleon).

\item $\pi N$ and $\eta N$ TDAs turn out to be interesting objects to study relying on  chiral dynamics and threshold soft pion theorems.

\item The Mellin moments of baryon-to-meson (and baryon-to-photon) TDAs in longitudinal momentum fractions
represent objects that are well suited to the lattice studies or with help of  the functional approaches based on the Dyson-Schwinger / Bethe-Salpeter equations.
\end{itemize}


\section{Next Steps}
\label{sec:next_step}

\subsection{A phenomenological search for the soft to hard transition}
\label{sec:soft_hard_transition}
Exclusive meson electroproduction above the resonance region, from near the photoproduction point to large $Q^2$, is a good handle to study the baryonic exchange reactions. The experimental and phenomenological efforts from JLab \cite{Park:2017irz, Li:2019xyp} raise further questions: What are forward-backward cross section ratios in other $u$-channel electroproduction interactions such as $\pi^0$, $\pi^\pm$, $\rho$, $\eta$, $\eta^\prime$ and $\phi$? Could the $t$-channel phenomenology recipe for mapping out the $W$ and $x$ dependence be applied to $u$-channel interactions? How would the $u$-channel interactions factorize (as illustrated in Fig.~\ref{fig:transition_2})? These important questions form the core bases for future studies.

The large acceptance, wide kinematics (in $Q^2$ and $W$) and forward tagging capability expected at the EIC provide a great opportunity to study near-forward and near-backward electroproduction of all mesons simultaneously. Combining the data collected at JLab 12 GeV and EIC, we aim to accomplish the following objectives to unveil the complete physics meaning of $u$-channel interactions:
\begin{itemize}

\item At the low $Q^2$ limit: $Q^2<2$ GeV$^2$, mapping out the $W$ dependence for electroproduction of all mesons at near-backward kinematics.

\item Extracting the $u$-dependence for the cross sections ($\sigma \propto e^{-b\cdot u}$) at a wide range of $Q^2$. This could be used to study the transition from a ``soft'' Regge-exchange type picture (where the transverse size of interaction is of order of the hadronic size) to the ``hard'' QCD regime.

\item Studying the model effectiveness from the hadronic Regge-based (exchanges of mesons and baryons) to the partonic description via Transition Distribution Amplitudes (exchanges of three quarks). This is equivalent to studying the non-perturbative to perturbative QCD transition for a given reaction.

\end{itemize}
\subsection{$u$-channel Regge Study Prospective}
\label{sec:$u$-channel regge_prospective}

The Reggeized nucleon exchange in the $u$-channel formulated in Eqn. (\ref{eqn:nucleon2}) has already been applied to $\omega$ photoproduction at backward angles \cite{clifft}. It can also be applied to the analysis of $\rho^0$ and $\phi$ vector meson cases to deepen our understanding of the role of the parton in the backward process.
\begin{figure}[ht]
\centering
\includegraphics[width=0.9\hsize]{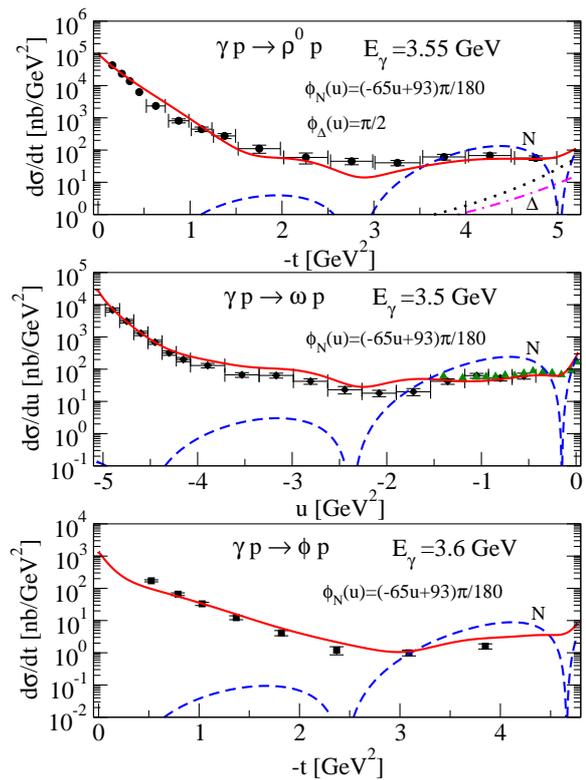}
\caption{ Differential cross sections for $\rho^0,\,\omega$ and $\phi$ photoproductions at similar energies $E_\gamma\simeq3.5$ GeV. The pure nucleon Reggeon is depicted by the dashed curve and the $\Delta$ Reggeon by dash-dotted one. The $\Delta$ with parton contributions
is denoted by dotted curve. Cross sections at forward angles are described by the Reggeized $t$-channel meson exchanges.  Parton contributions are considered via the nucleon isovector and isoscalar form factors with the relative phases $\phi(u)$ chosen to agree with data. Note that a single relative phase $\phi(u)=(-65u+93)\pi/180$ is applied to all the cases at the nucleon dip, $u\simeq-0.15$ GeV$^2$. Data are taken from Refs. \cite{battaglieri2001,clifft,anciant}. } 
\label{fig:regge6}
\end{figure}

\subsubsection{Backward vector-meson photoproduction}

Since backward $\phi(1020)$ photoproduction only allows the exchange of isoscalar component coupling to nucleon, as in the case of $\omega$, the Regge model calculations for these
two reactions are worth comparing within the same framework.
Moreover, the reaction $\gamma p\to\phi p$ will be beneficial for investigating the
flavor contents of quarks because the $s\bar{s}$ component is involved.
Figure \ref{fig:regge6} presents differential cross sections for photoproductions of lighter vector mesons, $\rho^0$, $\omega$ and $\phi$ at similar energies $E_\gamma\simeq3.5$~GeV. 
The Reggeized meson exchanges for $\sigma+\pi+f_2+$Pomeron \cite{yu5} could reproduce the cross section in the small $-t$ region to an extent.
The nucleon and parton contributions from the extended model in Eqn. (\ref{eqn:nucleon2}) offer good description to the data at backward angles with the $\phi NN$ coupling constants $g^v_{\phi NN}=3.2$ and $g^t_{\phi NN}=0$ chosen. 
Both reactions share the nucleon isoscalar form factor $F^{(s)}(u)$ in Eqn. (\ref{eqn:isoff}) parameterized by GPDs.
It should be noted that the data rising at $-t\approx3.9$ GeV$^2$
prohibit vanishing of the coupling constant $g^v_{\phi NN}$, and hence, cast a doubt on the validity of the OZI suppression in this reaction.

However, in contrast to these simple pictures, the study of parton contributions
to backward $\rho^0$ photoproduction
needs to include the exchange of the $\Delta^+$ 
in addition to proton exchange in the $u$-channel. 
As before, the $u$-channel nucleon Reggeon for the $\rho^0$ process is 
given by Eqn. (\ref{eqn:nucleon2}) with the coupling constants
$g^v_{\rho NN}=2.6$ and $g^t_{\rho NN}=9.62$, but in this case the nucleon isovector form factor is parameterized by the GPDs in Eqn. (\ref{eqn:quark}), i.e.,
\begin{eqnarray}
\hspace{2em}F_1^{(v)}(u)=\int_0^1dx\left[u_v(x)-d_v(x)\right]x^{-(1-x)\alpha'_qu}.
\end{eqnarray}

The $u$-channel Born term for the $\Delta$ exchange is a new entry to the photoproduction amplitude in Eqn. (\ref{eqn:nucleon2}) which is expressed as,
\begin{eqnarray}\label{eqn:delta1}
&&M_{u(\Delta)}={2\over3}\bar{u}(p')\Gamma^{\alpha}_{\gamma N\Delta}(k)
{\rlap{/}p'-\rlap{/}k+M_\Delta\over u-M_\Delta^2}\Pi^\Delta_{\alpha\beta}(p'-k)
\nonumber\\&&\hspace{1.5cm}\times
\Gamma^{\beta}_{\rho N\Delta}(q) u(p)
\end{eqnarray}
with the isospin coefficient $2/3$.
The $\gamma N\Delta$  and $\rho N\Delta$ vertices are given by
\begin{eqnarray}\label{eqn:ndelta}
&&\Gamma^{\alpha}_{\gamma N\Delta}(k)=ec_1\left(\rlap{/}\epsilon k^\alpha-\rlap{/}k\epsilon^\mu\right)\gamma_5\\
&&\Gamma^{\beta}_{\rho N\Delta}(q)
={f_{\rho N\Delta}\over m_\rho}
\gamma_5\left(\rlap{/}\eta q^\beta
-\rlap{/}q \eta^{\beta}\right)
\label{eqn:rhodelta}
\end{eqnarray}
where the coupling constant $c_1=2.38e$ GeV$^{-1}$ \cite{JONES19731} and $f_{\rho^0 p\Delta^+}=-5.05$ are chosen for the leading term and others are neglected for simplicity \cite{yu:2020}. The $\Delta$ exchange is in itself gauge invariant and the Reggeization is written as   
\begin{eqnarray}
\label{eqn:delta}
&&{\cal M}=
M_{u(\Delta)}\times\left(u-M_\Delta^2\right)
\nonumber\\&&\hspace{1cm}\times
\left({\cal R}^\Delta(s,u)+e^{i\phi(u)}F^{(v)}_\Delta(u)\widetilde{\cal
R}^\Delta(s,u)\right)\hspace{0.3cm}
\end{eqnarray}
for the extended model with the $\rho N\Delta$ transition 
form factor for the parton contributions.
The Regge propagator for the $\Delta$ is simply given by $J=3/2$, but the complex phase is chosen in Eqn. (\ref{eqn:regge}) in subsection 2.1. 

Thus, besides the GPDs for the isovector form factor
at the $\rho NN$ vertex,
the knowledge of the GPDs for the  $\rho N\Delta$ transition form factor
is further required. On the other hand, the GPDs for the $\rho N\Delta$ 
form factor can be constructed by using the $\rho NN$ vertex function
\cite{guidal}, as $\Delta$ and nucleon
are the same soliton states in the large $N_c$ 
limit \cite{frankfurt2000}. 
Thus, the isovector vector $N\to\Delta$ transition form factor relevant
to the $\rho N\Delta$ vertex in Eqn. (\ref{eqn:rhodelta}) can be constructed by the isovector magnetic form factor of nucleon, $F_2^p(u)-F_2^n(u)$, which is parameterized as,  
\begin{eqnarray}
\label{eqn:isoff_1}
&&F^{(v)}_\Delta(u)={G_M(0)\over\kappa_V}\int_{0}^{1}dx
\biggl[{e_u\kappa_u\over N_u}u_v(x)(1-x)^{\eta_u}
\nonumber\\&&\hspace{2.3cm}-{e_d\kappa_d\over N_d}(1-x)^{\eta_d}d_v(x)\biggr]
x^{-(1-x)\alpha'_qu}
\end{eqnarray}
with $\kappa_V=3.7$ and $G_M(0)\approx3.02$ taken from the
phenomenological value \cite{TIATOR2001205}.The slope $\alpha'_q=0.3$ GeV$^{-2}$ is chosen. $e_u$ and $e_d$ are
quark charges and $\eta_u=1.713$ and $\eta_d=0.566$ are parameters for the normalization of quark magnetic moments.
The trajectory $\alpha_\Delta(u)=0.9u-0.25$ is used for the $\Delta$
Reggeon ${\cal R}^\Delta$ and $\widetilde\alpha_\Delta(u)=0.9u+0.2$ for the $\widetilde{\cal R}^\Delta$.
The result of the reaction $\gamma p\to\rho^0 p$ at backward angles \cite{battaglieri2001} is presented in the upper panel of Fig. \ref{fig:regge6} 
to reveal the respective roles of proton and $\Delta^+$ Reggeon exchanges in the $u$-channel. The pure $\Delta^+$ contribution is denoted by the dash-dotted curve which is smaller than the nucleon by a factor of $10$. It is likely that the $u$-channel nucleon exchange dominates the backward photoproduction of these lighter vector mesons.


The JML model~\cite{Laget:2019} has already provided over the years an interpretation of these vector meson production channels. This section summarizes the results of the concurrent/parallel approach of B-G Yu.

\subsubsection{Backward $\pi^-\Delta^{++}$ photoproduction}

From the theoretical point of view, backward $\gamma p\to\pi^-\Delta^{++}$
reaction measured by the LAMP2 group \cite{BARBER1981135} is of significance
and should be paid attention to as well,
because it allows the single spin-3/2 $\Delta^{++}(1232)$ exchange only in the $u$-channel.
Therefore, the information concerning the GPDs for the
axial form factor at the $\pi N\Delta$ vertex can be obtained on the clean 
experimental background (similar to the GPDs
for the isoscalar form factor at the $\omega NN$ vertex in the
$\gamma p\to\omega p$ and $\gamma^* p\to\omega p$ reactions.)
As before, following the large $N_c$ argument \cite{frankfurt2000}, 
the GPDs for the $\pi NN$ form factor replaces those of $\pi N\Delta$ 
transition form factor with helicity-dependent parton distribution function (hPDF)
$\Delta q_v$ taken from Ref. \cite{florian}.

The issues concerning the reaction $\gamma p\to\pi^-\Delta^{++}$ 
are, firstly, the
determination of the slope and intercept of $\Delta(1232)$ trajectory
which are still uncertain, and secondly, the feasibility of GPDs for the
axial form factor at the $\pi N\Delta$ vertex.
Based on the suggested form of the $\pi N\Delta$ vertex with the GPD
taken only for the leading term for simplicity \cite{frankfurt2000}, 
we try to test whether the Regge model with parton contributions can reproduce the LAPM2 data.
Of course, this sort of attempt could be a guide for further study of 
hPDF for the $N\Delta$ excitation.

Within the framework of Sec.~\ref{sec:Regge}, the photoproduction amplitude is given by Eqn. (\ref{eqn:delta}) which now replaces $M_{u(\Delta)}$ with  $\left(M_{s(p)}+M_{t(\pi)}+M_{u(\Delta)}+M_c\right)$ 
for charge conservation of the $u$-channel $\Delta^{++}$ exchange. It requires proton and pion exchanges with the contact interaction term further in Eqn. (\ref{eqn:delta}) \cite{yu3}.
Given the Regge propagator for the $\Delta$ as before,
the axial form factor $F^{(a)}_\Delta(u)$ is introduced to 
the $\pi N\Delta$ vertex with parton contributions. 
The conventional model corresponds to the one
without the second term  in Eqn. (\ref{eqn:delta}).
For the analysis of LAMP2 data in the energy range $E_\gamma=3.5$ - 4.4 GeV,
we consider both the spin-3/2 projection $\Pi^{\mu\nu}_\Delta$ and the vertex
$\gamma \Delta\Delta$ fully as in Ref. \cite{yu4}, but not
employ the minimal gauge prescription.
The intercept of $\Delta$ trajectory is, in general, taken
to be non-negative in most Regge model calculations. However, it would be
advantageous to explain empirical data by using the intercept
chosen as a negative number, i.e., $\alpha_\Delta(u)=0.9u-0.25$, in the Reggeized
Born term model as presented in Eqn. (\ref{eqn:delta}).

The helicity-dependent GPDs
for the axial form factor at the $\pi N\Delta$ vertex
can be expressed as 
\begin{eqnarray}\label{eqn:gpdaxff}
\hspace{2em}&&F^{(a)}_\Delta(u)={1\over1.24}\int_0^1dx\left[\widetilde{H}^u_v(x,u)-\widetilde{H}^d_v(x,u)\right]
\end{eqnarray}
for the valence quark and
\begin{eqnarray}\label{eqn:hgpds}
\hspace{4em}\widetilde{H}^q_v(x,u)=\Delta q(x)f_q(x,u)
\end{eqnarray}
with the polarized parton distribution for
valence quarks taken from Ref. \cite{florian}
\begin{eqnarray}
\Delta u_v=&&\nonumber\\
&&\hspace{-3em}0.677x^{-0.308}(1-x)^{3.34}(1-2.18x^{0.5}+15.87x),\nonumber\\
\Delta d_v=&&\nonumber\\
&&\hspace{-3em}-0.015x^{-0.836}(1-x)^{3.89}(1+22.4x^{0.5}+98.94x)
\end{eqnarray}
respectively. The ansatz for the profile function is
chosen as
\begin{eqnarray}
\label{eqn:ansatz}
\hspace{6em}f_q(x,u)=x^{-(1-x)\alpha'_qu},  
\end{eqnarray}
with $\alpha'_q=0.3$ GeV$^{-2}$ in 
Eqn. (\ref{eqn:hgpds}) above.

\begin{figure}[ht]
\centering
\includegraphics[width=0.9\hsize]{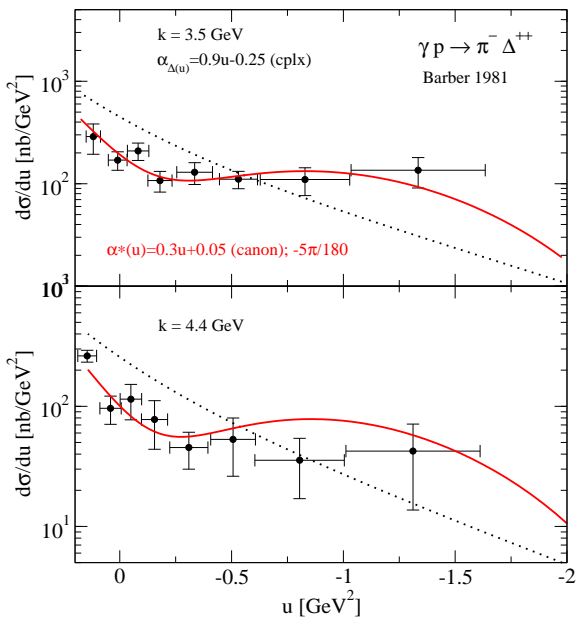}
\caption{ Differential cross sections for $\gamma p\to\pi^-\Delta^{++}$ from LAMP2 data at $E_\gamma=3.5$ and 4.4 GeV. Dotted curves show the pure hadronic contributions from the conventional Regge model with the complex phase for the $\Delta^{++}$ Reggeon. (Red) solid curves are the results of the extended Regge model  with the GPDs for the axial form factor at $\pi N\Delta$ vertex included in the parton contribution. The relative phase $\phi(u)=-5\pi/180$ together with canonical phase and $\widetilde\alpha=0.3u+0.05$ are taken for the $\widetilde{R}^\Delta$. Data are taken from Ref. \cite{BARBER1981135}.
} \label{fig:regge7}
\end{figure}

\begin{figure}[ht]
\centering
\includegraphics[width=0.9\hsize]{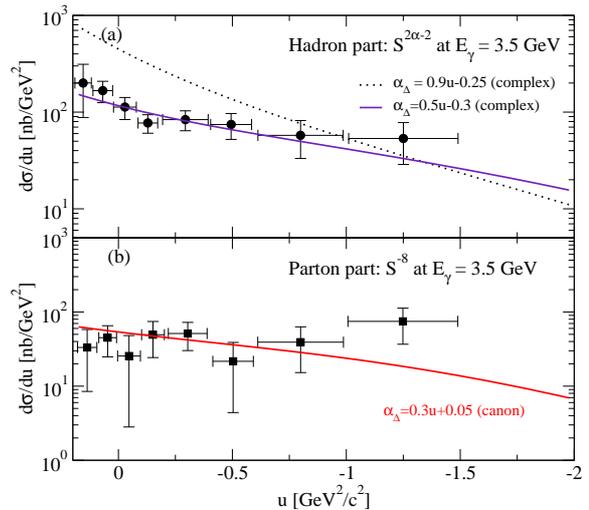}
\caption{Hadron and partonic parts of differential cross section for $\gamma p\to\pi^-\Delta^{++}$ at $E_\gamma=3.5$ GeV within the same framework as in Fig. \ref{fig:regge2}. Data are taken from Ref. \cite{BARBER1981135}.} \label{fig:regge5}
\end{figure}

In Fig. \ref{fig:regge7}, differential cross sections at $E_\gamma=3.5$
and 4.4 GeV are presented, respectively. The dotted curve represents the 
cross section from the conventional model prediction and 
the cross section including parton contributions is depicted by 
the solid curve. In both models, the complex phase is chosen for 
the $\Delta$ Reggeon ${\cal R}^\Delta$ in order to be consistent 
with the apparent absence of the dip of the $\Delta$ trajectory, 
which should appear at $u=-1.38$ GeV$^{-2}$, otherwise. 
Although there exists the model-dependence in the current approach, 
the LAMP2 data cannot be described from the conventional point of view without parton contributions.

Hadron and parton contributions in the separated form from each other 
are shown in Fig. \ref{fig:regge5}. In panel (a) the dotted curve 
in Fig. \ref{fig:regge7} resumes a large deviation from the data
of the hadron part. 
As illustrated by the solid curve, 
the data in actual needs the $\Delta$ trajectory to be 
$\alpha_\Delta=0.5u-0.3$, for instance, which can hardly be accepted 
in the usual sense.
The data from the parton contribution to the cross section in the lower panel (b)  
shows the $s^{-8}$ scaling predicted by the quark counting rule for the 
photoproduction of hadron \cite{collins84}. Therefore, such a slope of the 
scaled cross section has to be reproduced by nearly vanishing slope of 
the trajectory $\widetilde\alpha_\Delta=0.3u+0.05$ of the modified Reggeon 
$\widetilde{R}^\Delta$ with the canonical phase. To summarize, the solid curve in Fig. \ref{fig:regge7} results from the sum of the 
dotted curve in (a) and the solid one in (b) via the adjustment 
of the relative phase $\phi(u)=-5\pi/180$ in Fig. \ref{fig:regge5}, though the independent part of the hadronic cross section requires the $\Delta$ trajectory unacceptable .

\subsubsection{Backward $\pi^+$ electroproduction}

In parallel with the JML approach, the final ``Regge'' based study  will be the backward reaction $\gamma^*p\to\pi^+n$ \cite{Park:2017irz}  from JLab 6~GeV, as analyzed in Fig. \ref{fig:CLAS_cross_section_u_channel}.  Within the current framework in Eqn. (\ref{eqn:nucleon2}) in Sec. \ref{sec:Regge} the Reggeized $\Delta^0$ exchange is written as Eqn. (\ref{eqn:delta}) in which case the $\rho N\Delta$ vertex is replaced by the $\pi N\Delta$ in Eqn. (\ref{eqn:delta1}) with the meson-baryon form factor from GPDs given in Eqn. (\ref{eqn:gpdaxff}). It is commonly used for the $\pi NN$ vertex in the neutron exchange in the $u$-channel as well. 
Given the $\gamma N\Delta$ vertex in Eqn. (\ref{eqn:ndelta}) the dipole form factor $F_2^n(Q^2)$ in Eqn. (\ref{eqn:dipoleff}) is taken for the $\Delta^0$ at $\gamma^*N\Delta$ as well as the neutron at $\gamma^*NN$ vertices with the cutoff mass $\Lambda=1.55$ GeV in common. The $\Delta^0$ Reggeon plays a role in the backward $\pi$ electroproduction, second to neutron. 
The $Q^2$-dependences of $\sigma_T$ and $\sigma_L$ decrease much steeper than those of $\omega$ electroproduction as shown in the right panel of Fig. \ref{fig:regge2}.

\subsection{Backward TCS and DVCS
\label{sec:btcs}
}

\begin{figure} [htb]
\centering
  \includegraphics[width=0.22\textwidth]{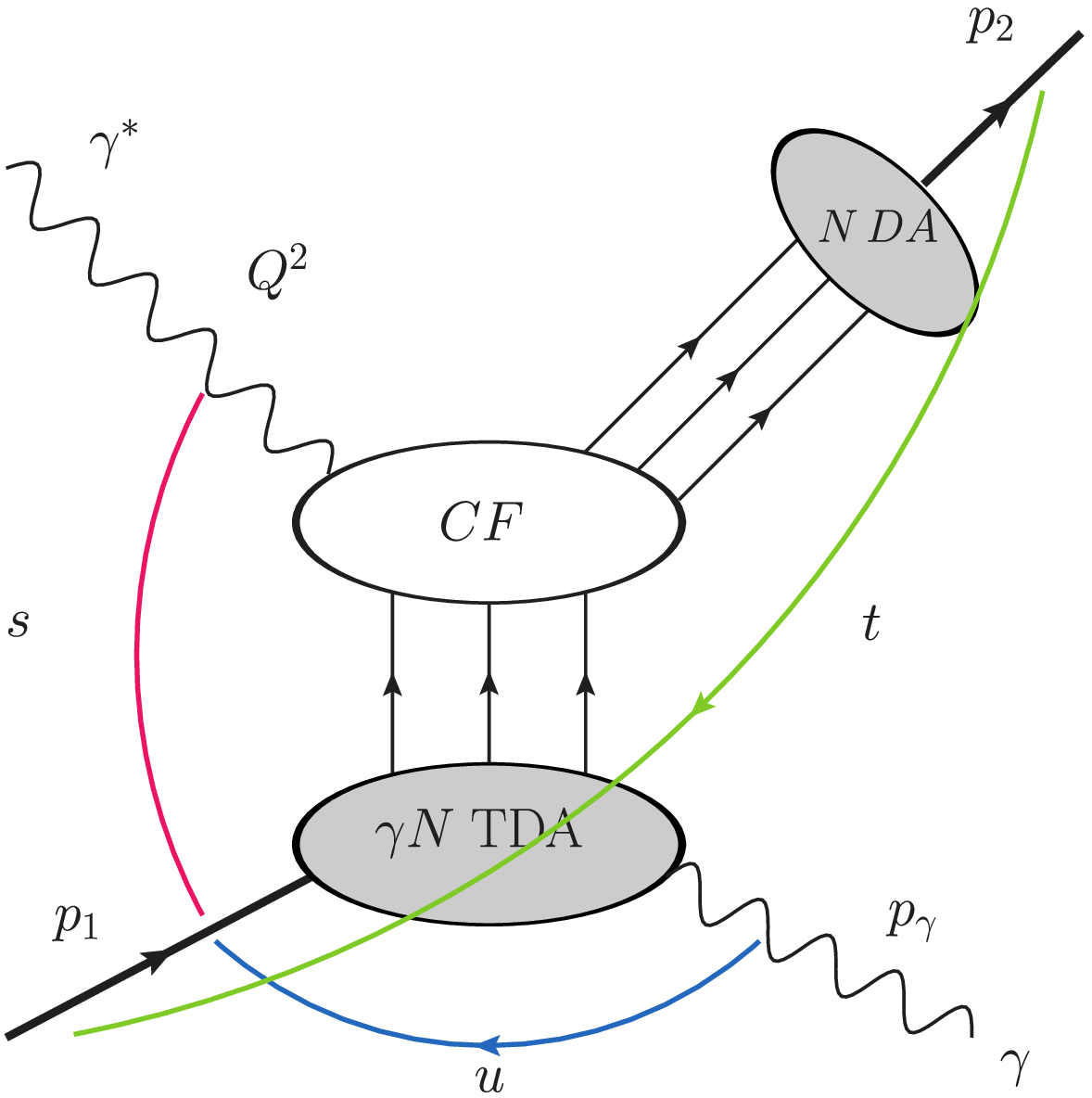}~~~~
  \includegraphics[width=0.23\textwidth]{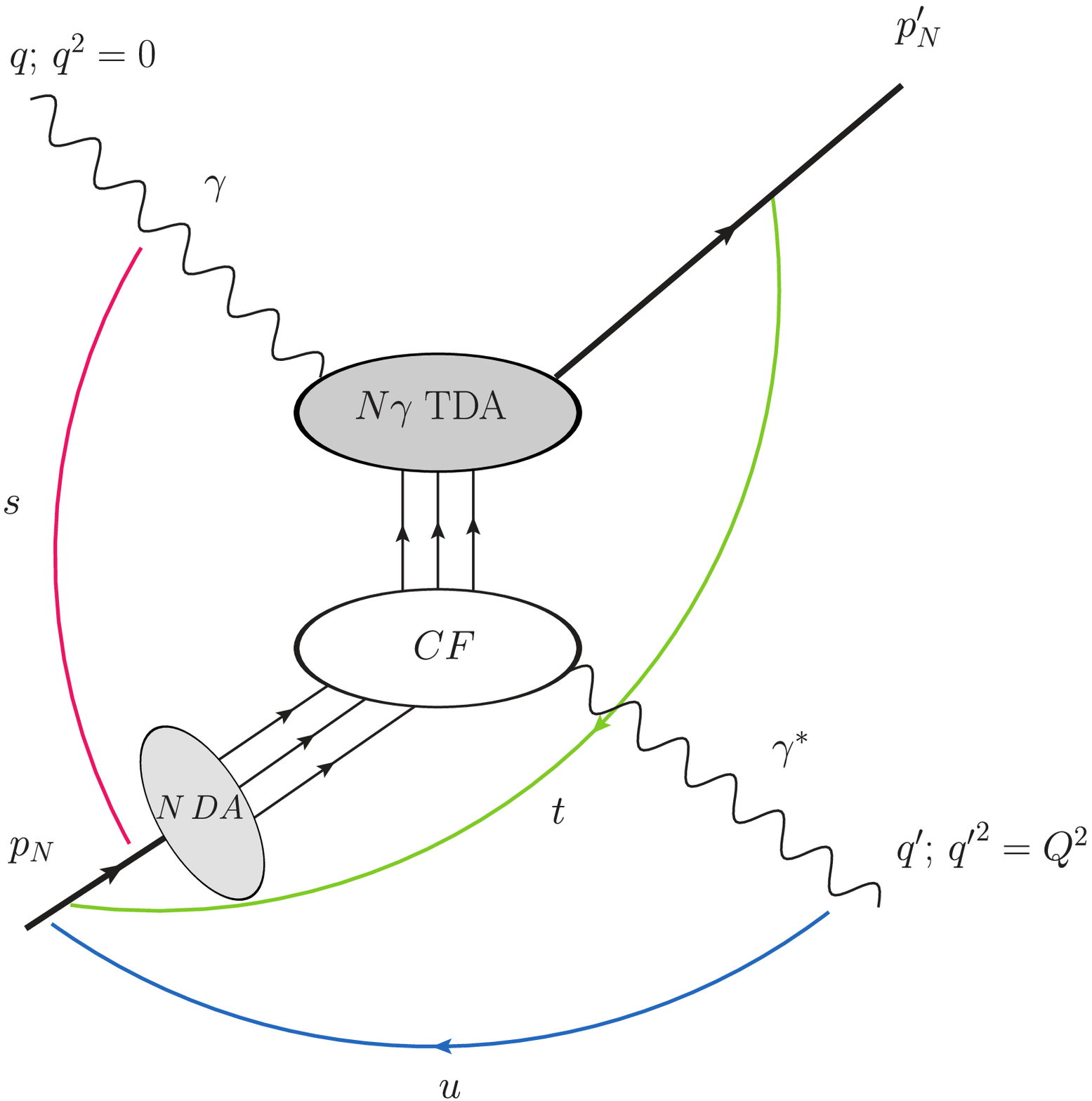}
\caption{Backward DVCS (left panel) and backward TCS (right panel) probe the same $N\to \gamma$ TDAs}
\label{fig:bTCS} 
\end{figure}
The study of 
DVCS in the forward region has been instrumental in the development of the collinear QCD description of deep exclusive reactions in terms of GPDs. The numerous data obtained at various energies, at HERMES, HERA, JLab and COMPASS were successfully analyzed in this framework (for a recent review, see e.g. \cite{Kumericki:2016ehc}). The peculiar feature of this reaction is that the Born term of the hard subprocess is a pure QED process. This property is shared with the exclusive photoproduction of a lepton pair, named time-like Compton scattering (TCS) \cite{Berger:2001xd}, in reference with the time-like nature of the virtual photon carrying the large scale needed for a perturbative expansion of the coefficient function. Both of these processes mix with pure QED Bethe-Heitler processes, which are dominant in most kinematical regions but serve as a magnifier thanks to the specific properties of the interference cross-section; for instance, the charge-exchange property of the produced lepton pair in TCS and in the QED process allows one to separate the interference cross section through a charge-odd observable. 

The $N\to\gamma$ and $\gamma \to N$ transition distribution amplitudes, which respectively enter the backward DVCS and TCS amplitudes, are very similar objects, as were the $N\to\pi$ and $\pi\to N$ TDAs discussed in the case of backward production of a $J/\psi$ with a pion beam \cite{Pire:2016gut}, the relation reading
\begin{equation}
\hspace{2em}TDA^{\gamma \to N} (x_i,\xi, u) = TDA^{N \to \gamma} (-x_i,-\xi, u)\,
\end{equation}
As for nucleon to meson TDAs, nucleon to photon TDAs are Fourier transformed  matrix element of the
\begin{equation}
\hspace{2em}\widehat{O}_{\rho \tau \chi}^{uud} =\varepsilon_{c_1 c_2 c_3} 
 u^{c_1}_{\rho}(\lambda_1 n)
u^{c_2}_{\tau}(\lambda_2 n)d^{c_3}_{\chi}(\lambda_3 n) \nonumber
\end{equation}
 operator, decomposed on leading twist Dirac structures as 
\begin{eqnarray}
&&
4 {\cal F} \langle \gamma(q, \epsilon_\gamma) | \widehat{O}_{\rho \tau \chi}^{uud}(\lambda_1n,\lambda_2n,\lambda_3n)|N^p(p_N,s_N) \rangle
\nonumber \\ &&
\hspace{2em} = \delta(x_1+x_2+x_3-2\xi)  \times m_N \times \\
&&\Big[
\sum_{\Upsilon= 1 {\cal E}, 1T, \atop 2 {\cal E}, 2T  } (v^{\gamma N}_\Upsilon)_{\rho \tau, \, \chi} V_{\Upsilon}^{\gamma N}(x_1,x_2,x_3, \xi, \Delta^2; \, \mu^2)
\nonumber \\ &&
+\sum_{\Upsilon= 1 {\cal E}, 1T,  \atop 2 {\cal E}, 2T  } (a^{\gamma N}_\Upsilon)_{\rho \tau, \, \chi} A_{\Upsilon}^{\gamma N}(x_1,x_2,x_3, \xi, \Delta^2; \, \mu^2)\nonumber\\&&
+
\sum_{\Upsilon= 1 {\cal E}, 1T,    2{\cal E}, 2T, \atop 3 {\cal E}, 3T,  4 {\cal E}, 4T } (t^{\gamma N}_\Upsilon)_{\rho \tau, \, \chi} T_{\Upsilon}^{\gamma N}(x_1,x_2,x_3, \xi, \Delta^2; \, \mu^2)
\Big]\,
\nonumber 
\label{eqn:VN_TDAs_param}
\end{eqnarray}
Because the photon has two polarization states, there are twice as many GPDs as for the pseudoscalar meson case, i.e. 16 leading twist GPDs. Four of them give a non-vanishing contribution to the amplitude at zero $\Delta_T$, which correspond to the helicity conserving amplitudes
\begin{equation}
\hspace{5em} T_{+-,-}^{+ \to +}, T_{ - +,-}^{+ \to +}, T_{- -,+}^{+ \to +}, T_{+ +,+}^{+ \to -} \,
\end{equation}
with the obvious notation $T_{\lambda_u \lambda_u,\lambda_d}^{\lambda_N \to \lambda_\gamma}$. The relation between TDAs and light-front helicity amplitudes reads at zero $\Delta_T$:
\begin{eqnarray}
\hspace{1em}V_{1 {\cal E}}^{ \gamma p}&= \frac{1}{2^{1 / 4} \sqrt{1+\xi}\left(P^{+}\right)^{3 / 2}} \frac{1}{m_N}\left[T_{+ -,-}^{+\to +}+T_{- +,-}^{+ \to +}\right]; \nonumber\\
A_{1 {\cal E}}^{ \gamma p}&=- \frac{1}{2^{1 / 4} \sqrt{1+\xi}\left(P^{+}\right)^{3 / 2}}  \frac{1}{m_N}\left[T_{+ -,-}^{+\to +}-T_{- +,-}^{+ \to +}\right];\\
T_{1 {\cal E}}^{ \gamma p}&=-\frac{1}{2^{1 / 4} \sqrt{1+\xi}\left(P^{+}\right)^{3 / 2}} \frac{1}{m_{N}}\left[T_{- -,+}^{+ \to +}+T_{+ +,+}^{+ \to -}\right];\nonumber \\
T_{2 {\cal E}}^{ \gamma p}&=-\frac{1}{2^{1 / 4} \sqrt{1+\xi}\left(P^{+}\right)^{3 / 2}} \frac{1}{m_{N}}\left[T_{- -,+}^{+ \to +}-T_{+ +,+}^{+ \to -}\right].\nonumber
\end{eqnarray}
Nucleon to photon TDAs thus allow to access new physics information on density probabilities for quark helicity configurations when a proton emits a photon. For instance the ratio 
\begin{equation}
\hspace{8em}\frac {\vert V_{1  {\cal E}}^{ \gamma p}\vert ^2+ \vert A_{1  {\cal E}}^{ \gamma p}\vert ^2}{\vert T_{1  {\cal E}}^{ \gamma p}\vert ^2+ \vert T_{2  {\cal E}}^{\gamma p}\vert ^2}
\end{equation}
gives access to the ratio $\frac{d^{h^u =-h^{u'}}(x_i)}{d^{h^u =+h^{u'}}(x_i)}$
which may be interpreted as the answer to the question: {\em Is the nucleon brighter when u-quarks have equal helicities}?

Counting the $\Delta_T$ factors in the Dirac structures accompanying the TDAs allows to get access to the  orbital angular momentum contribution to nucleon spin. For instance, since  the spinorial structure attached to $T_{4 {\cal E}}^{ \gamma p}$ contains $\Delta_T^{3}$, which implies  $L=3$, the 
 $T_{4 {\cal E}}^{ \gamma p}$ TDA measures the helicity amplitude $T_{- -,-}^{+ \to -}$, and the ratio 
 \begin{equation}
\hspace{5em} \frac {\vert T_{4 {\cal E}}^{ \gamma p}\vert ^2}{\vert V_{1  {\cal E}}^{ \gamma p}\vert ^2+ \vert A_{1  {\cal E}}^{ \gamma p}\vert ^2+\vert T_{1  {\cal E}}^{ \gamma p}\vert ^2+ \vert T_{2  {\cal E}}^{ \gamma p}\vert ^2} 
 \end{equation}
 measures the ratio of density probabilities for three units vs zero unit of orbital angular momentum between the three quarks when a proton emits a photon.


As for the nucleon to meson TDAs case, there is an impact picture of nucleon to photon TDAs: Fourier transforming to impact parameter : $\Delta_T \to b_T $ allows to access the question: 
{\em Where in the transverse plane does the nucleon emit a photon} ?

 Phenomenologically complete models for the proton to photon TDAs need to be constructed. A VDM framework may allow to relate them to nucleon to transversely polarized vector meson TDAs. The phenomenology of the reactions involving these new TDAs remains to be worked out.
The case for backward DVCS was briefly discussed in \cite{Pire:2004ie,Lansberg:2006uh} and the case for backward TCS is under current investigation. The experimental difficulties are quite different to access these two processes, but their theoretical and physical contents are very similar. Indeed their hard amplitudes are equal (up to a complex conjugation) at Born order, and differ only in a controllable way at NLO  \cite{Muller:2012yq}. Moreover, the QED process which shows a pronounced peak in the forward region is quite harmless in the backward region.

\subsection{Regge model of the forward-backward asymmetry in the $\pi^-\eta^{(\prime)}$ production}
\label{sec:forward-backward_asymmetry}

\begin{figure}
\centering
\includegraphics[width=0.33\textwidth]{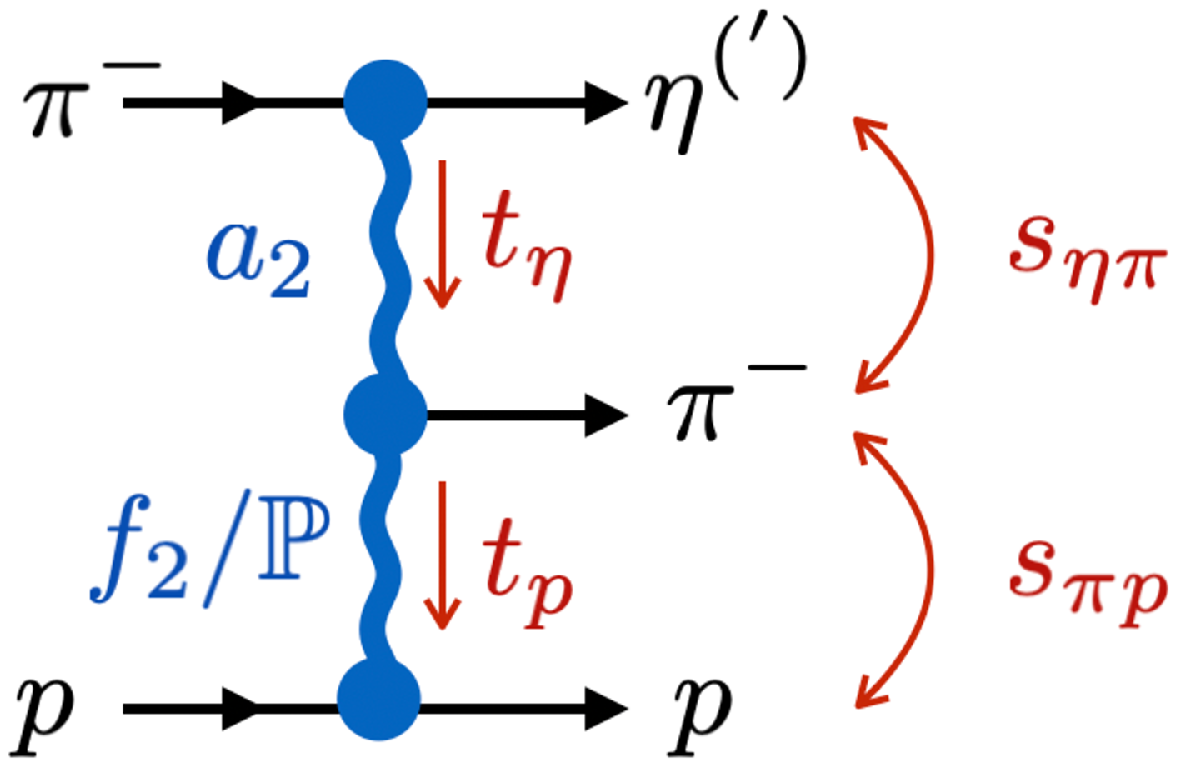} \\
\includegraphics[width=0.33\textwidth]{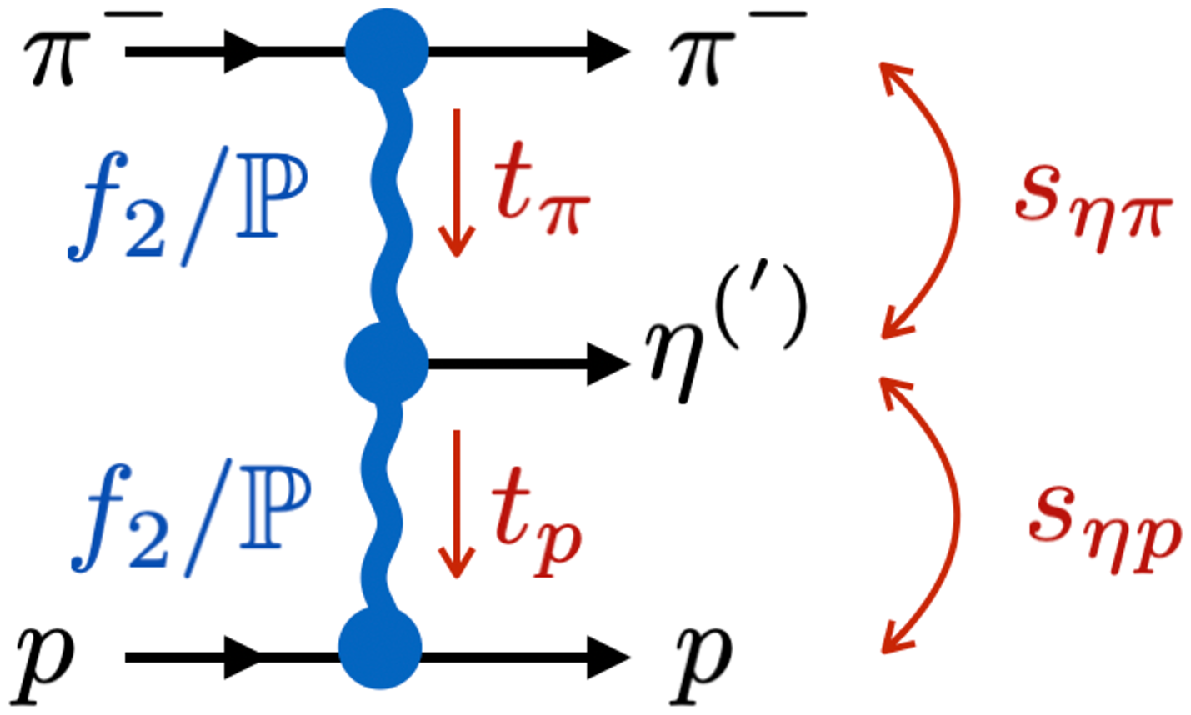}
\caption{Fast-$\eta$ (top) and fast-$\pi$ (bottom) amplitudes.}
\label{fig:exchanges}
\end{figure}

The $\pi^-p\to \pi^-\eta^{(\prime)}p$ reaction has been recently studied by the COMPASS collaboration at CERN \cite{Adolph:2014rpp}. The analysis has revealed that in the high energy region the $\pi\eta^{(\prime)}$, system is produced in two kinematic regimes. In these regimes the laboratory frame direction of the $\eta^{(\prime)}$ meson is either forward or backward. The Gottfried-Jackson frame analysis of the $\eta^{(')}$ polar angle distribution thus reveals a characteristic forward-backward asymmetry, with a rapidity gap in-between, see Fig. 2 in \cite{Adolph:2014rpp}. In terms of the polar angle intensity this asymmetry can be defined as
\begin{subequations}
\begin{align}
\hspace{4em}A(m ) \equiv & \,   \frac{F(m ) - B(m )}{F(m ) + B(m ) } \, ,
\label{eq:intensities_asy}\\
F(m ) \equiv & \int_0^1 \diff\cos \theta \,
I_\theta(m ,\cos\theta) \, , \label{eq:intensities_fwd} \\
B(m ) \equiv & \int_{-1}^0 \diff\cos \theta \, I_\theta(m ,\cos\theta) \, , \label{eq:intensities_bwd}
\end{align} \label{eq:fbintensities}
\end{subequations}
with $F(m)$ and $B(m)$ being the forward and backward intensities.
It was shown in \cite{Bibrzycki:2021rwh} that this angular dependence can be described in terms of the double Regge exchange depicted in Fig.\ref{fig:exchanges} where the upper reggeon is exchanged in either the $t$- or $u$-channel.   

The general form of the double Regge amplitude expressed in terms of channel invariants reads \cite{ShimadaMartinIrving1978}
\begin{align}
\label{sigamp}
    T=&-K \Gamma(1-\alpha_1)\Gamma(1-\alpha_2)\\ \nonumber
    &\biggl[
    (\alpha' s)^{\alpha_1-1}(\alpha' s_2)^{\alpha_2-\alpha_1}\xi_1\xi_{21}\hat{V}_1+ \\ 
    &\hspace{0.5em}(\alpha' s)^{\alpha_2-1}(\alpha' s_1)^{\alpha_1-\alpha_2}\xi_2\xi_{12}\hat{V}_2
    \biggl]\nonumber
\end{align}
%
where signature factors $\xi_i$, $\xi_{ij}$, the $\hat{V}_i$ functions and the $K$ factor are defined in \cite{Bibrzycki:2021rwh}.

\begin{figure*}[h!]
    \begin{tabular}{cc}    
    \includegraphics[width=0.45\textwidth]{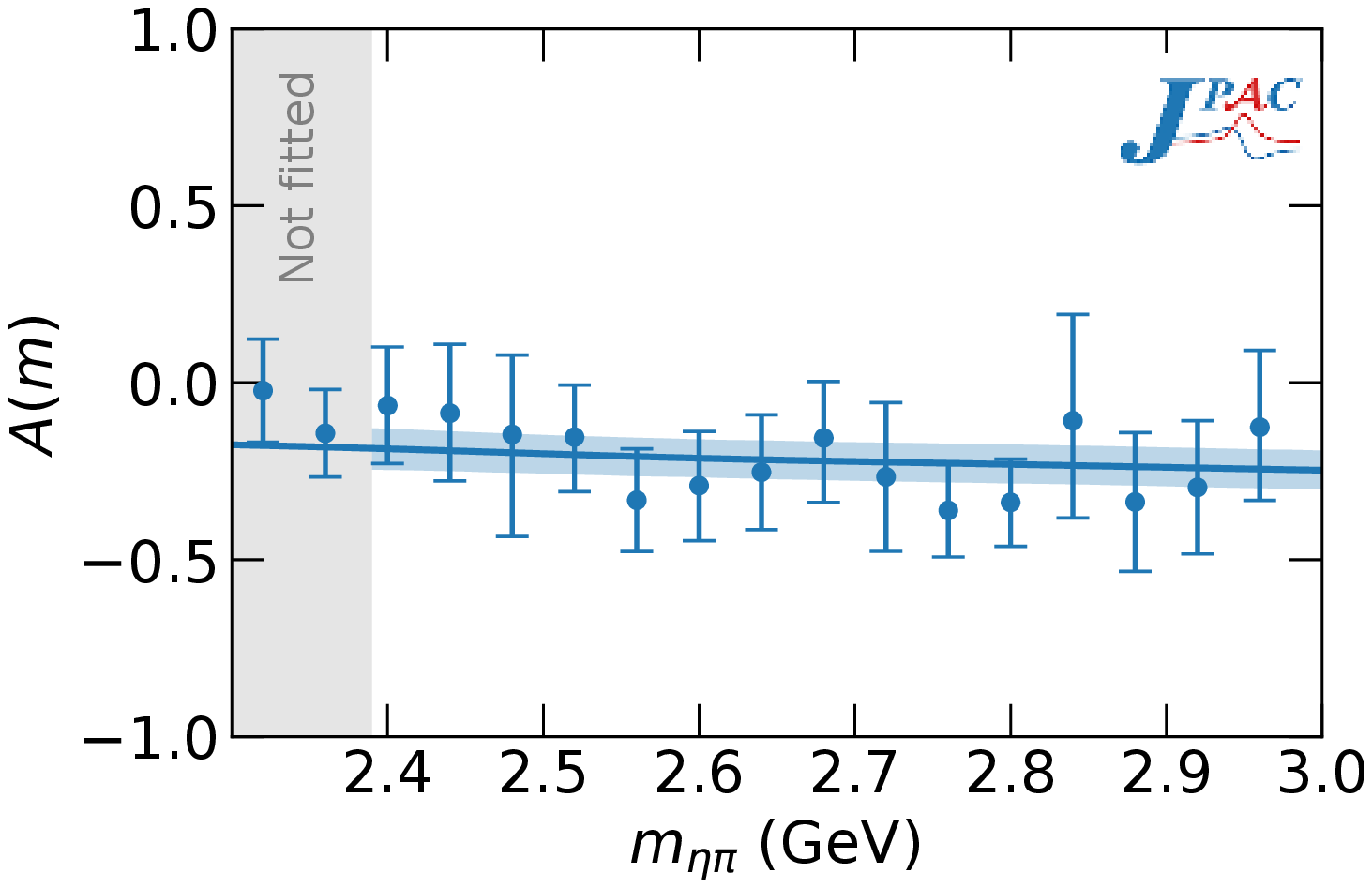} &
    \includegraphics[width=0.45\textwidth]{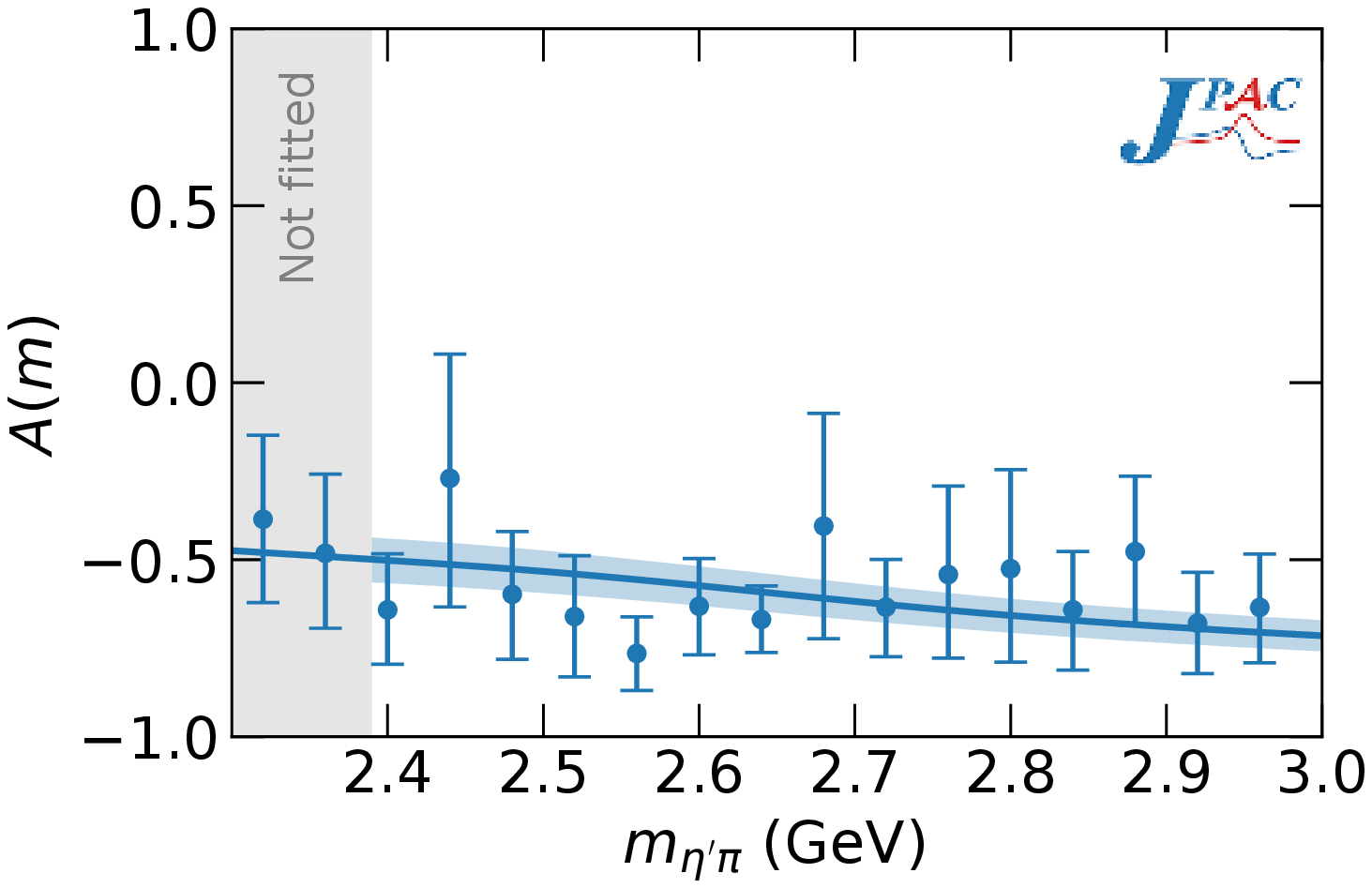}
    \end{tabular}
      \caption{Forward-backward intensity asymmetry as defined in Eq.~\eqref{eq:fbintensities} for $\eta \pi$ (left) and $\eta' \pi$ (right) from Ref.~\cite{Bibrzycki:2021rwh}.
      }
      \label{fig:asy.fit}
\end{figure*}

In this model both $\alpha_1$ and $\alpha_2$ correspond to $2^{++}$ exchanges as represented in Fig. \ref{fig:exchanges}. 
In the high $\eta^{(')}\pi$ mass region, the asymmetry $A(m)$ originates from the asymmetry between the fast $\eta^{(')}$ and fast $\pi$ diffractive production depicted in Fig.~\ref{fig:exchanges}. In the low mass $\eta^{(')}\pi$ mass region, where $\eta^{(')}\pi$ resonances are produced, the forward-backward asymmetry originate from the interference between even and odd waves. As odd partial waves in the $\pi\eta^{(')}$ system are exotic \cite{KLEMPT20071,PhysRevLett.122.042002}, the high energy forward-backward asymmetry $A(m)$ can be formally related to the production of exotics resonance through special dispersion relations, also known as the Finite Energy Sum Rules \cite{Dolen:1967jr,Mathieu:2015gxa}.
In particular, the $P-$wave that brings the largest contribution to the forward-backward asymmetry can be related to the lightest hybrid resonance $\pi_1$. The forward-backward asymmetries predicted by the model for the $\pi\eta$ and $\pi\eta'$ channels and compared with experimental data are shown in Fig.\ref{fig:asy.fit}.  
As the double Regge exchange mechanism is largely independent of particular trajectories and final states, one can expect the similar phenomena to appear in the high energy photoproduction experiments like CLAS12 and GlueX at JLab. In particular reliable description of the $\pi\eta^{(\prime)}$  photoproduction may be decisive for the observation of hybrid mesons in electromagnetic processes. 

\subsection{Upcoming experimental program and opportunities} 
\subsubsection{Upcoming JLab Hall C Measurements}
\label{sec:hallc_12gev}

\begin{figure} 
\centering
  \includegraphics[width=0.52\textwidth]{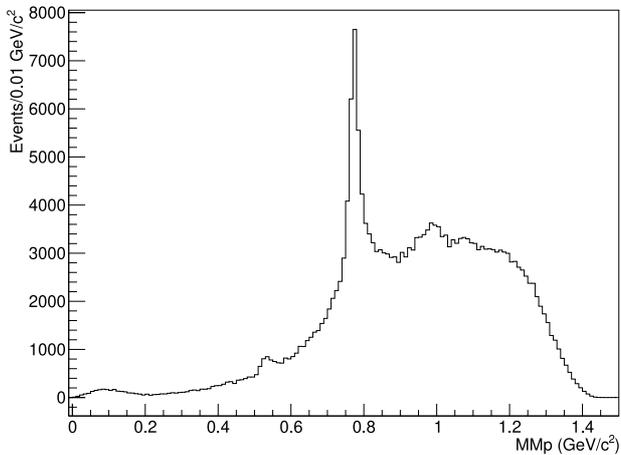}
\caption{Preliminary missing mass spectrum for electron-proton coincidence events identified in the measurement of the~$Q^{2}=3$~GeV$^{2}$,~$W=2.32$~GeV,~high $\epsilon$ (0.88) central angle setting of the KaonLT experiment. A strong $\omega$ peak, and lesser $\eta$, $\phi$ peaks, are evident.}
\label{fig:KLT_HighEps} 
\end{figure}

Similar to the 6~GeV era Hall~C measurements presented in Sec.~\ref{sec:hallc_omega}, additional $u$-channel meson electroproduction data were fortuitously acquired during the KaonLT experiment (E12-09-011) \cite{E12-09-011}.  The primary purpose for the acquisition of these data is the study of the $K^+$ electromagnetic form factor, but the detector apparatus allows $p(e,e'p)X$ data to be acquired in parallel.  Data were taken well above the resonance region ($W=2.32$--3.02 GeV), at selected settings between $Q^2=0.5$ and 5.50 GeV$^2$.  For each $Q^2, W$ setting, data were taken at two beam energies, corresponding to $\Delta\epsilon\sim 0.25$, so that L/T/LT/TT separations can be performed. 

In the KaonLT experiment, the recoil electron is detected in the HMS spectrometer, and the forward-going hadron is detected in the SHMS 11~GeV/c spectrometer.  The hadron identity is determined with the use of heavy gas ($C_4F_{10}$) and aerogel Cherenkov detectors at the SHMS focal plane, the coincidence time difference between the two spectrometers, and the time difference between the SHMS and accelerator RF pulses.  This allows the high momentum, forward-going proton data in the SHMS to be cleanly identified and analyzed.  For all of these data, $u\sim 0$.  
Fig.~\ref{fig:KLT_HighEps} shows an example missing mass spectrum acquired in 2018--19, at high $\epsilon$ for a single setting at $Q^2=$3.0 GeV$^2$, $W=2.32$ GeV.

A prominent $\omega$ peak, corresponding to production with very low lab momentum, is observed.  Preliminary indications are that this cross section is dominantly transverse. $u$-channel $\phi$ production is particularly interesting to study, as it is uniquely sensitive to the $s \overline{s}$ content of the nucleon.  Finally, for some settings, small $\eta$ and $\eta'$ peaks are observed at the edges of the coincidence missing mass acceptance.
In 2021--2022, additional $u$-channel data are expected to be acquired during the PionLT experiment \cite{E12-19-006}, up to $Q^2$=8.5~GeV$^2$, contributing further to the extensive L/T-separated data set expected from Hall~C as the data are acquired and analyzed.  TDA model cross section predictions for the PionLT settings are available in Ref.~\cite{Pire:2015kxa}.

\begin{figure} 
\centering
  \includegraphics[width=0.48\textwidth]{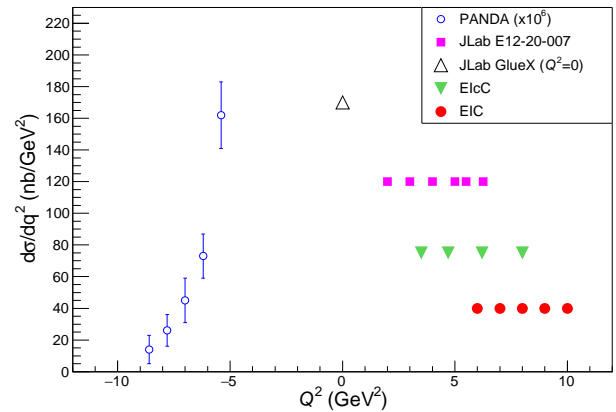}
\caption{The anticipated global data set of $d\sigma/dq^2 (\gamma^*p \rightarrow p\pi^0)$ vs $Q^2$ at fixed $s=10$ GeV$^2$. Projected results from $\overline{\textrm{P}}$ANDA (FAIR) $q^2$ ($-Q^2$) scaling are in open blue circle; projected JLab E12-20-007 measurements are in magenta square; projected EIC measurements are in red full circle; JLab GlueX photoproduction measurement ($Q^2 = 0$ GeV$^2$) is indicated by the open triangle. These experimental programs are elaborated in the relevant subsections.}
\label{fig:u_channel_pi0} 
\end{figure}

JLab experiment E12-20-007 \cite{E12-20-007} aims to further study the TDA framework by probing the $^1$H$(e, e^{\prime}p)\pi^0$ exclusive electroproduction reaction
over the $2<Q^2<6.25$~GeV$^2$ kinematic range, at fixed $W=3.1$~GeV ($s=10$~GeV$^2$) and $-u_{min}$. This is the first dedicated $u$-channel experiment approved by JLab, rather than using data acquired fortuitously in a separate measurement. 
The experiment will utilize the 11 GeV $e$ beam on an unpolarized liquid hydrogen target (LH$_2$), in combination with the high precision HMS, SHMS spectrometers available in Hall~C. The key observable involves the detection of the scattered electrons in coincidence with energetic recoiled protons, and resolving the exclusive $\pi^0$ events using the missing mass reconstruction technique (Eqn.~\ref{eqn:missmass}). The separated cross sections, $\sigma_{T}$ , $\sigma_{L}$, and the $\sigma_{T}/\sigma_{L}$ ratio at 2-5 GeV$^2$, will directly challenge the two predictions of the TDA model, $\sigma_T = 1/Q^8$ and $\sigma_T \gg \sigma_L$, in $u$-channel kinematics. This will be an important step forward in validating the existence of a backward factorization scheme and establishing its applicable kinematics range.

The right panel of Fig.~\ref{fig:u_channel_pi0} illustrates a prospective $Q^2$ evolution study ($-10<Q^2<10$ GeV$^2$), combining backward ($u \sim u_{\textrm {min}}$) exclusive $\pi^0$ production data from JLab, $\overline{\textrm{P}}$ANDA and EIC, at fixed $W=10$ GeV. A preliminary study has confirmed the feasibility of studying $e+p\rightarrow e'+p'+\pi^0$ over the range $6.25 < Q^2 < 10.0$~GeV$^2$. The EIC offers unique opportunity to provide a definitive test of TDA predictions beyond JLab 12 GeV kinematics. Furthermore, the EIC result is anticipated to play a significant role in the extraction of TDAs.

\subsubsection{Upcoming JLab CLAS12 Measurements}
\label{sec:CLAS12}

Due to its large acceptance, CLAS12 can be used to map out electroproduction cross sections and asymmetries over a wide range of kinematics, covering the forward and backward regimes simultaneously in a $Q^{2}$ range up to 8 GeV$^{2}$. In addition, CLAS12 can provide detailed measurements of the azimuthal dependence of the cross section. The CLAS12 data, which are approved and partly already recorded from other experiments, will be used to extract cross sections and asymmetries for different mesons in the backward regime and compare them to measurements in the forward regime. Hard exclusive production of $\pi^{+}$, $\rho$ and $\phi$, in addition to other channels, will be studied with CLAS12 in backward kinematics.


Regarding the detection of the backward produced meson, the estimated $u$-channel acceptances of CLAS12 are as follows: 
\begin{itemize}
\item {$\pi^0$: good acceptance up to $-t$ of 5-6 GeV$^2$. $u$-channel measurements not possible.}
\item {$\pi^+$: nearly full coverage of $t$ and $u$ acceptance. Measurements up to $-u_{min}$.}
\item {$\rho/\omega$: the $\pi^{\pm}$ final states from $\rho\to\pi^+\pi^-$ and $\omega\to\pi^+\pi^-\pi^0$ decay channels can be well measured with full coverage of the $t$ and $u$ acceptance.  In principle, a measurement up to $-u_{min}$ is possible, but the threshold on each of the pions introduces some limitations at small $-u$ compared to the $\pi^+$ channel. Nevertheless, one can get close to $-u_{min}$.} 
\item {$\phi$: The $K^+K^-$ channel can be well measured with full coverage of the $t$ and $u$ acceptance, but statistics will be very limited at small $-u$. Similar limitations due to thresholds at very small $-u$ as for $\rho$/$\omega$.}
\end{itemize}

\subsubsection{Meson photoproduction at GlueX}
\label{sec:GlueX_CLAS12}

The GlueX experiment~\cite{adhikari:21}, in Jefferson Lab Hall D, features a linearly-polarized 9~GeV real photon beam delivered to a large acceptance detector system. The collaboration recently completed its first phase of running, and analysis efforts of this data set are well underway. Thanks to its near $4\pi$ detector converge in photon detection, extraction of exclusive meson differential cross sections are accessible at both forward and backward kinematics, thus offering a full $t$ or $u$ coverage.

\begin{figure} 
\centering
  \includegraphics[width=0.49\textwidth]{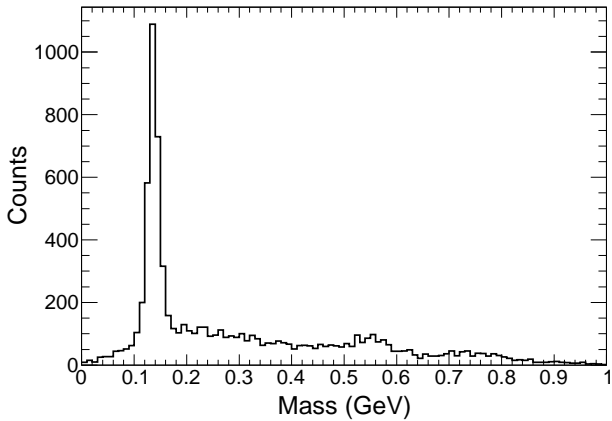} 
  \caption{Invariant mass of two photons in the exclusive $\gamma p \rightarrow \gamma\gamma p$ reaction with beam energy $E_\gamma =$5.4~GeV and $-t>$3 GeV$^2$ to select the $u$-channel dominated regime.}
\label{fig:gluex_pi0_mass} 
\end{figure}

\begin{figure} 
\centering
  \includegraphics[width=0.45\textwidth]{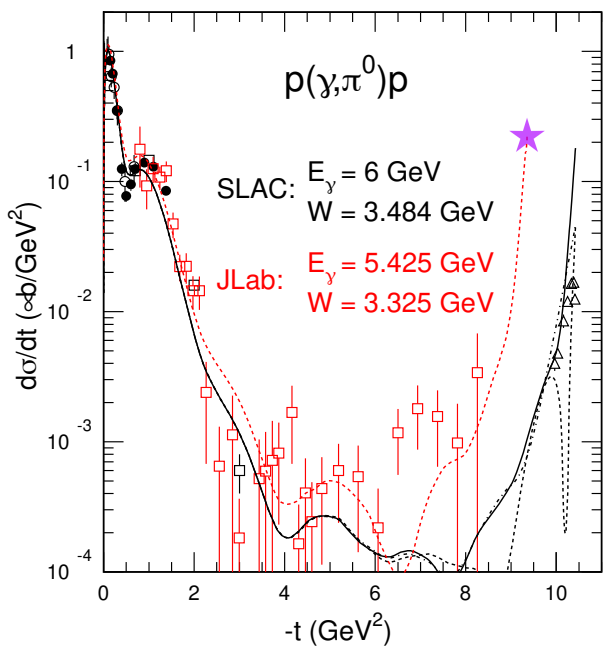} 
  \caption{The differential cross-section of the $p(\gamma, \pi^0)p$ reaction at $E_\gamma\sim5.45$~GeV over the full angular range. The red dashed line curve is the JML model prediction at $E_\gamma= 5.425$ GeV. Open triangles represent the measurement from Ref.~\cite{Tompkins:1965}, filled circles from ~\cite{Anderson:1970}, empty squares from ~\cite{Kunkel:2018}. Full line: full JML model prediction at $E_\gamma = 6$ GeV; the dot-dashed: $\Delta$ Regge pole exchange contribution; dashed line: nucleon Regge pole and unitary cuts. The purple star represents the impact of GlueX $\pi^0$ data at a photon energy of 5.4 GeV at $u\sim u_{min}$. Note that the GlueX data contain full $-t$ coverage. These predictions of the JML model were first published in ~\cite{Laget:2021}.}
\label{fig:gluex_pi0} 
\end{figure}

\begin{figure} 
\centering
  \includegraphics[width=0.49\textwidth]{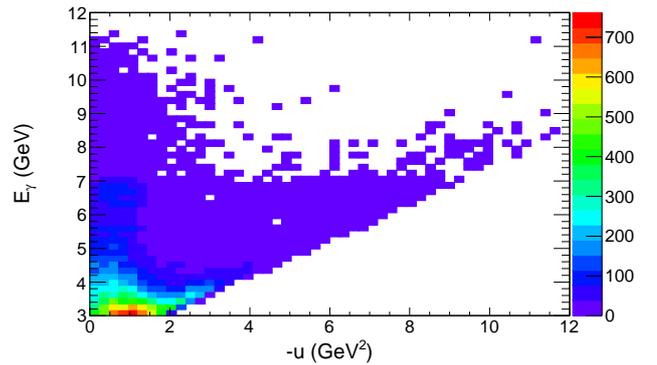} 
  \caption{$E_\gamma$ vs $-u$ coverage of the GlueX experiment for exclusive $\pi^0$ photoproduction with $-t>$3 GeV$^2$ to select the $u$-channel dominated regime. 
  }
\label{fig:gluex_pi0_phasespace} 
\end{figure}

In $u$-channel kinematics ($u \sim u_{min}$), the final state protons absorb the momentum from the photon probe, and recoil forward towards the Forward-Drift-Chamber and Time-of-Flight (TOF) detectors. The time difference between the Start-Counter (which surrounds the target cell) and TOF is an effective criterion in selecting such proton events. The produced $\pi^0$ is emitted at a wide-angle with low momentum, and the decayed photons ($\pi^0\rightarrow\gamma\gamma$) are captured by the Barrel Calorimeter. An example of a reconstructed $\pi^0\rightarrow \gamma\gamma$ event peak is shown in Fig.~\ref{fig:gluex_pi0_mass}, at a photon energy of $E_\gamma=5.4$ GeV. To ensure the exclusivity of the $\gamma p \rightarrow p \pi^{0}$ final state, a missing mass cut, $-0.01 < m_{miss}^2 <$ 0.01 GeV$^2$, is applied.

Figure~\ref{fig:gluex_pi0} shows prior measurements of the $\pi^0$ photoproduction differential cross-section as a function of $-t$, compared with theoretical predictions from the JML model~\cite{Laget:2021}. 
Here, the Regge exchange diagrams are similar to those in Fig.~\ref{fig:omega_t_slope}. Although the JML model predicted a rise in differential cross-section in $u$-channel kinematics (for $-t\rightarrow t_{max}$), there has been no direct experimental measurement confirming the existence of the anticipated $u$-channel cross-section peak for $E_\gamma \sim 5.4$ GeV.

Figure~\ref{fig:gluex_pi0_phasespace}, shows the $E_{\gamma}$ vs $-u$ coverage from a portion of the GlueX experimental data collected in Fall 2018, with beam energy coverage from $ 3 < E_{\gamma} < 11 $~GeV.  
With this large kinematics coverage, one could:
\begin{itemize}
\item extract the $u$-dependence of $\pi^0$ cross section in the range: 4 $< E_{\gamma}< $ 10 GeV. 
\item verify the $W$ scaling of the $\pi^0$ cross section near $u_{min}$ follows the expected $(W^2-m_p^2 )^{-2}$ dependence. 
\end{itemize}

These $u$-channel photoproduction studies can be extended to other mesons, including $\eta$, $\omega$ and $\phi$. The theoretical prospective for these studies was described in Sec.~\ref{sec:$u$-channel regge_prospective}.  In the case of $\omega$ photoproduction, the previous experimental data covered the $E_\gamma$ range: $E_{\gamma} < 5$~GeV~\cite{yu} (see Fig.~\ref{fig:regge6}), and GlueX can extend the data coverage up to $E_\gamma=11$ GeV with high statistics and precision. It is also worth mentioning that the lower energy $\omega$ photoproduction data have shown a slight `dip' feature at $u~-0.2$ GeV$^2$, (see Fig.~\ref{fig:regge6} middle plot), and the location of the `dip' was linked to nucleon structure by the Regge approach in Sec.~\ref{sec:Regge}. In GlueX, this feature can be carefully studied at higher $E_\gamma$ energy.
 
\subsection{Backward production in ultra-peripheral collisions}
\label{sec:UPC}

High-energy photonuclear interactions may be studied using ultra-peripheral collisions (UPCs) of heavy ions at RHIC and the LHC \cite{Klein:2020fmr,Baltz:2007kq}.  One nucleus emits a photon, while remaining intact.  This photon then interacts with the other nucleus.  This leads to reactions such as coherent or incoherent photoproduction of vector mesons ($V$), $AA\rightarrow AAV$.  In $pA$ collisions, the photon usually comes from the ion, with the proton serving as target.  Here, we will consider backward production in UPCs, focusing on $pA$ with the proton as target. 

In UPCs, the nuclei do not interact hadronically (in simple terms, the impact parameter $b$ is larger than twice the nuclear radius $R_A$), but interact electromagnetically.  The slightly-virtual ($Q^2 < (\hbar/R_A)^2$) photons are from the nuclear electromagnetic fields.  
The photon flux scales with the nuclear charge as $Z^2$, so heavy nuclei produce high photon fluxes.  UPCs have been used to study a variety of physics, including studies of two-photon physics, measurements of gluon shadowing, and nuclear imaging \cite{Klein:2019qfb}.  UPCs can probe energies far beyond those that are accessible at fixed target facilities, reaching center of mass energies $W_{\gamma p}$ above 1~TeV at the LHC. Because of the high photon fluxes, large light-meson data samples can be collected, with one STAR analysis using 470,000 events, even after tight cuts \cite{Klein:2018grn}. With the upcoming LHC Run 3, large samples of light and heavy vector mesons are expected, with CMS, ALICE and LHCb all expected to accumulate more than 1 million $J/\psi$ events \cite{Citron:2018lsq}.  
UPCs can probe proton targets, using $pp$ or $pA$ collisions (here $A$ implies nuclei with atomic number greater than one), or heavy ion targets in $AA$ collisions. In $pp$ or $AA$ collisions, there is uncertainty over which nucleus emitted the photon and which was a target, leading to destructive interference between the two possibilities in certain kinematic regions \cite{Klein:1999gv, Abelev:2008ew}. After combining the two possibilities, $d\sigma/dy$ becomes symmetric, and there is an ambiguous relationship between rapidity and photon energy.

Here, we will focus on $pA$ interactions, $pA\rightarrow VpA$. $pp$ collisions could also be used to study backward production on proton targets, but the backgrounds are likely to be larger, and they also suffer from the bidirectional interference. 
Backward production on ions in $AA$ collisions should also be very interesting, since it involves neutron targets.  However, since the target nucleon will be ejected from the ion, coherent production will likely not be possible and the cross-sections will be lower than for coherent production. Furthermore, the luminosities for $AA$ collisions are lower than for $pA$ collisions.  Thus, $pA$ collisions are likely to be the best initial venue for backward production studies. 

To study the rates and kinematics of backward production in UPCs, we adopt a simple paradigm: that backward production is very similar to the usual forward vector meson production (also $pA\rightarrow VpA$, with $t$ usually small)
except that $t$ and $u$ are swapped.
The cross-section for forward vector meson production may be modelled as \cite{Klein:1999qj}
\begin{equation}
\hspace{5em}    \frac{d\sigma}{dt} = (XW_{\gamma p}^\epsilon + YW_{\gamma p}^{-\eta})\cdot e^{-Bt},
    \label{eqn:vmsigma}
\end{equation}
where $X$, $Y$, $\epsilon$ and $\eta$ are constants that depend on the meson being produced. The slope $B$ may also depend on the final state, but much more weakly; it represents the squared size of the production region.  The term $XW_{\gamma p}^\epsilon$ accounts for Pomeron exchange, while $YW_{\gamma p}^{-\eta}$
backward production is similar to Reggeon production, involving the exchange of non-zero quantum numbers, with a cross-section that decreases with increasing $W$.  Because of the decreasing $\gamma-p$ cross-section with increasing photon energy, most of the Reggeon-exchange production is concentrated at large $|y|$ \cite{Klein:2019avl}.  

Here, we focus on $\omega$ production via Reggeon exchange (neglecting the Pomeron exchange component, which is more important at large $W_{\gamma p}$), because it is the most studied vector meson in backward production.  For the $\omega$, $Y= 180 \mu$b/GeV$^2$ and $\eta=1.92$, and $B\approx 12$ GeV$^{-2}$ for a proton target.
With the model, the cross-section for backward production is 
\begin{equation}
\hspace{5em}     \frac{d\sigma}{du} = (AW_{\gamma p}^{-\eta})\cdot e^{-Cu},
    \label{eqn:vmsigmabackward}
\end{equation}
where a fit to fixed-target photoproduction data found $A=4.4 \mu$b/GeV$^2$ and $\eta=2.7$.  We also take $C=32$~GeV$^{-2}$.  This is larger than the $B$ found for forward production. This difference is expected in Regge theory, since backward production involves baryon trajectories, while forward production involves meson trajectories.

\begin{figure} 
\centering
  \includegraphics[width=0.5\textwidth]{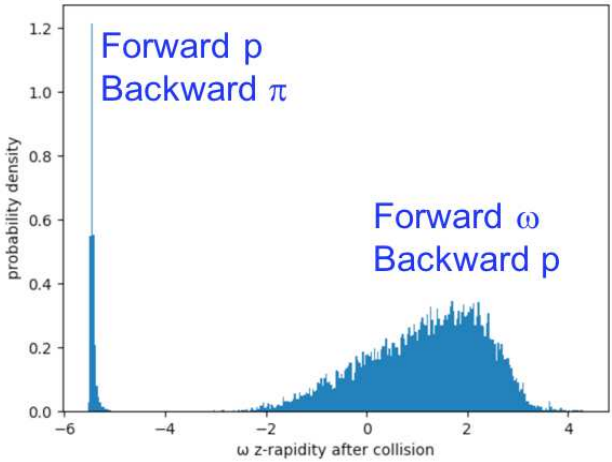} 
  \caption{Rapidity (labelled $z$) distribution for forward and backward photoproduced $\omega$ at RHIC with proton targets.  The calculation is done within the framework of the STARlight Monte Carlo~\cite{Klein:2016yzr}.  For forward production, the $\omega$ is produced in the region $-3<y< 5$, while the proton is at $-6<y<-5$.  For backward production, the two rapidity ranges are reversed.  In this figure, the proton target is coming from the right ($+y$ direction). }
\label{fig:UPComega} 
\end{figure}


$\eta$ for backward production is somewhat larger than for forward production, so the cross-section is more concentrated near threshold. Threshold production, with $W_{\gamma p}=M_p+M_\omega$, occurs for a photon energy $k=1.44$~GeV in the target frame, or $k' = k/2\gamma$ in the lab frame, Here, $\gamma\gg 1$ is the Lorentz boost of the target ion, so the lab frame photon energies are very low.  One can relate the target-frame photon energy to the final state rapidity, with 
\begin{equation}
\hspace{5em}     k'= M_V e^{(\pm y)}.
\end{equation}
The $\pm$ sign is due to the two-fold ambiguity as to which ion is the photon emitter.  For $pA$, the ion usually emits the photon, and the sign of the rapidity depends on the beam convention.   Because the cross-section is largest near-threshold, the bulk of the production occurs with the $\omega p$ center of mass at large $|y|$; $|y|$ increases with increasing beam energy, with $|y|=5.7$ for 250 GeV protons at RHIC, and $|y|=4.75$ for 7~TeV protons at the LHC.  Figure \ref{fig:UPComega} shows the calculated rapidity distributions for backward and forward production of $\omega$ at RHIC.  The events with $|y|<4$ are from the usual forward production, while the peak around $y\approx -5.7$ is for backward production; this is nearly equal to the beam rapidity.  The large $|y|$ is unfortunately beyond the reach of the existing STAR detector, and the in-construction sPHENIX. The peak moves to larger $|y|$ with increasing beam energy, with $|y|\approx 9$ with 7~TeV protons at the LHC.  A RHIC run at lower collision energy would move the backward production toward mid-rapidity.  A run with 41 GeV protons would put the peak at $|y|\approx 3.85$.

The other `observability' question involves rates.  By comparing Eqns. \ref{eqn:vmsigma} and \ref{eqn:vmsigmabackward}, it is clear that the overall backward production rate is a few percent of the forward production rate.  Since the forward production cross-sections are large, rate is unlikely to be a limiting factor for observing backward production.  Instead, good acceptance at large $|y|$ is the key.

It is worth exploring various possibilities for observing backward production at large $|y|$.  The proposed STAR forward upgrade will provide tracking and calorimetery \cite{Shi:2020gex} in the pseudorapidity range $2.5<\eta<4$.  This is a wide enough range to observe $\omega$ photoproduction, as long as the target proton beam energy is reduced to below about 50 GeV.   For the $\omega$, the most promising final state might be $\pi^0 \gamma$, since it relies only on calorimetry. Or, one could focus on tracking-based detection and search for backward production of $\rho\rightarrow\pi^+\pi^-$.  The production characteristics are likely to be similar as for the $\omega$ channel.  The rates for backward production of $\rho$ are poorly known, but the forward rate is about 10 times higher than for the $\omega$.
It might also be interesting to investigate mesons with significantly different masses.  Lighter mesons may be produced more copiously, but will be at larger $|y|$.  $\pi^0\rightarrow\gamma\gamma$ is of interest as a simple state, and the cross-section is likely to be high.  The cost is the larger production rapidity.
Alternately, one could consider significantly heavier mesons, such as charmonium. 


\subsection{$u$-channel Meson Electroproduction at the EIC}
\label{sec:EIC}

For $u$-channel exclusive meson electroproduction processes, the higher center of mass energy at the EIC offers an unique opportunity for further extend the kinematics coverage. Photoproduction and electroproduction of the $\omega$ and $\pi^0$ seem to be the most experimentally accessible.

While detection problems for 
backward meson production at the EIC are in many respects similar to those in UPCs, the different photon source -- an electron, instead of a heavy ion, changes 
the photon spectrum, 
allowing photons with large $Q^2$ to become usable as hard probes of the hadronic process. 
In fact, the backward production part of Fig. \ref{fig:UPComega} is applicable, with a small rapidity adjustment to account for the slightly higher proton beam energy.  
That said, the EIC detectors are much more attractive for backward production studies, because they 
should have a much larger angular acceptance, with central detectors that instrument out to $|\eta|<4$ and forward detectors that cover most of the forward region; the latter are important for observing backward meson production. 

We can use the same model presented in Section \ref{sec:UPC} to roughly estimate the cross-sections and rates for $u$-channel production at the EIC, with the new ingredient of $Q^2$ dependence.  Here, we assume that the $Q^2$ dependence is the same for forward and backward production, and use the parameterization for the $\omega$ in eSTARlight \cite{Lomnitz:2018juf}:
\begin{equation}
    \sigma_{\gamma^*p\rightarrow \omega p}(W,Q^2) =   \sigma_{\gamma^*p\rightarrow \omega p}(W,Q^2=0 ) \big(\frac{M_\omega^2}{M_\omega^2+Q^2}\big)^n\,,
    \label{eq:Q2}
\end{equation}
where $n=c_1+c_2(Q^2+M_V^2)$, and $c_1$ and $c_2$ are taken from Ref. \cite{Aaron:2009xp}.  As with UPCs, the $u$-channel production cross-section for the $\omega$ is very roughly of order 1\% of the forward production cross-section.  Since the expected forward production rate for the $\omega$ is more than a billion events/$10^7$ seconds, it can be seen that the backward production rate is ample, as long as the final states are discernable in the detector.  Assuming that the $Q^2$ scaling in Eq. \eqref{eq:Q2} holds for backward production, the electroproduction rate should be over 10 million/$10^7$~s run period. 

\begin{figure} 
\centering
  \includegraphics[width=0.5\textwidth]{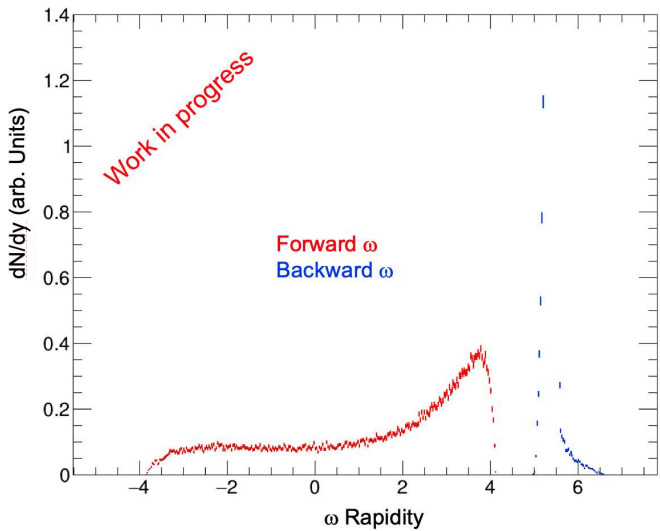} 
  \caption{Rapidity distribution for forward and backward photoproduced $\omega$ at the EIC for 18~GeV electrons colliding with 100~GeV/n protons. The calculation is done within the framework of the eSTARlight Monte Carlo \cite{Lomnitz:2018juf}.  The red curve shows the rapidity distribution of normally (forward) produced $\omega$, while the blue curve is for $\omega$ $u$-channel production.  At the same time, the red curve shows the proton final state rapidity for $\omega$ backward production, while the blue curve shows the protons for conventional $\omega$ photoproduction. This study covers all of the available $\gamma$-$p$ invariant mass range.}
\label{fig:EIComega} 
\end{figure}

Fig. \ref{fig:EIComega} shows the expected $\omega$ rapidity distribution for backward and forward $\omega$ production, for 18~GeV electrons incident on 100~GeV/n protons. The scattered protons are well within the rapidity acceptance of the EIC reference design \cite{AbdulKhalek:2021gbh}, but, for photoproduction ($Q^2\approx 0$), they will have very low $p_T$, so may not always be detectable.  For electroproduction, the protons should be more visible. 

In view of the desirability of observing final states containing charged particles, it is worth considering ways to shift the meson rapidity.  One approach is to reduce the proton beam energy; this will shift the $\omega$ peak to smaller $y$ and also broaden it somewhat.  With a 41~GeV proton beam, and a final state proton at rapidity 0, the typical $\omega$ rapidity is 4.6 -- still out of the acceptance of the central detector.  However, with the same beam energy, if the proton is scattered toward forward rapidities, the typical $\omega$ rapidity will be smaller, and might be visible in the central detector.

The electron tagging system coverage extends to low $Q^2$ (corresponding to $\eta$ coverage up to 6.9).  The low expected event rate, combined with high precision for the tagging system, is sufficient to support meson photoproduction studies ($Q^2\sim 0$ GeV$^2$).  However, due to the kinematic limitations of the electron tagging system, the scattered electron is not always detected. The loss of kinematic constraints may lead to an increase of background.  

The $\omega$ are well into the far-forward region (along the initial proton beam direction), where a Zero Degree Calorimeter (ZDC) can be used to tag the neutral particle fragments, so that decays like $\omega\rightarrow \pi^0\gamma$ may be the most promising. Unfortunately, the paucity of charged particle detection reduces the number of mesons that can be studied; the $\rho$, which is likely to be the most copiously produced meson, decays almost exclusively to $\pi^+\pi^-$.   The $\phi$, which is of great interest because it shares no quark flavors with the incident proton, has only a 1.3\% branching ratio to an all-neutral state, $\eta\gamma$, which is followed by $\eta\rightarrow\gamma\gamma$ or the harder to reconstruct $\eta\rightarrow 3\pi^0$.

In the $\pi^0$ electroproduction sector, the impact of EIC data is illustrated in Fig.~\ref{fig:u_channel_pi0}.  It shows a prospective $Q^2$ ($10 < Q^2 < 10$ GeV$^2$) evolution, combining backward ($u \sim u_{\textrm {min}}$) exclusive $\pi^0$ production data from JLab E12-20-007, $\overline{\textrm{P}}$ANDA, and EIC, at fixed $s=10$ GeV$^2$. 
A preliminary study has confirmed the feasibility of studying the  $e+p\rightarrow e'+p'+\pi^0$ interaction in the range: $6.0 < Q^2 < 10.0$~GeV$^2$. A data set combining E12-20-007 (Sec.~\ref{sec:hallc_12gev}) and EIC for exclusive $\pi^0$ production will offer a definitive challenge to the $1/Q^{10}$ scaling prediction of the TDA formalism (see Sec.~\ref{sec:TDA}).


\begin{figure} 
\centering
  \includegraphics[width=0.49\textwidth]{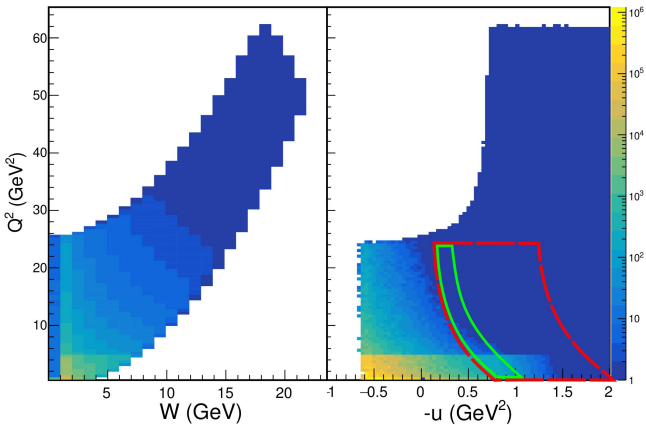}
  \caption{Left: $Q^2$ vs $W$ coverage for $e+p\rightarrow e^\prime+ p^\prime + \pi^0$, where at least one $\gamma$ is detected by the ZDC. Right: $Q^2$ vs $-u$ coverage for all available $s$ values, $0<s<400$ GeV$^2$. The red enclosed region represents $Q^2$ vs $-u$ coverage at $9<s<11$ GeV$^2$; the green enclosed region is a subset of the red, and presents the coverage of events with both photons detected by the Zero-Degree Calorimeter.} 
\label{fig:u_channel_phasespace} 
\end{figure}

The study of $e+p\rightarrow e^\prime+ p^\prime + \pi^0$ at $-u_{min}$ arises surprisingly naturally, thanks to the 4$\pi$ coverage of the EIC detector package and forward-tagging capability. A feasibility study has shown the optimal collision energy option: 5 GeV electron beam on a 100 GeV proton~\cite{AbdulKhalek:2021gbh}, for fixed $s=10$ GeV$^2$ at $u\sim u_{min}$. The corresponding available $Q^2$ vs $-u$ phasespace is shown in the right panel of Fig.~\ref{fig:u_channel_phasespace}.
At the kinematic range of interest, the scattered electrons at pseudorapidity $|\eta|<1.5$ and $p_{e}\sim 5.4$~GeV/c will be detected by the Electron-End-Cap, well within the EIC specification
~\cite{Accardi:2012qut, EIC:RDHandbook, AbdulKhalek:2021gbh}; the Zero Degree Calorimeter (ZDC) will be used to detect decayed $\pi^0\rightarrow\gamma\gamma$ for momenta from 40 to 60 GeV/c; for the forward recoiled proton, the detector and material studies show the Hadron-End-Cap will provide a silicon tracker to cover $\eta$ range up to 3.5: $|\eta|<3.5$. One would need a dedicated detector to tag the recoiled proton at $\eta \sim 4.1$ at $\phi = 180^\circ$, otherwise the missing mass reconstruction technique will be applied to resolve the proton. Note that the feasibility of the missing mass reconstruction technique remains to be demonstrated at the EIC.

To extract the differential cross section of the exclusive $\pi^0$ events, the event selections include the following scenarios:
\begin{itemize}

\item{All final state particles are detected, including $e^{\prime}$, $p^{\prime}$ and 2$\gamma$. A feasibility study~\cite{AbdulKhalek:2021gbh} projected 20 to 30\% double $\gamma$ detection efficiency for $\pi^0$ at 40 to 60 GeV/c, respectively. Here, $p_\pi=$40 GeV/c corresponds to $Q^2 \sim 10$ GeV$^2$.  It is also worth noting the hit pattern of the two photons forms a ring around the high occupancy spectator neutron region at the ZDC plane from other tagged diffractive processes.} 

\item{$e^\prime$, $p^\prime$ and a single $\gamma$ (from decayed $\pi^0$) are detected. The lost photon will likely be consumed by the steering magnet arrays upstream of the ZDC. Under this scenario, one could rely on a detailed simulation of the known physics backgrounds, such as $u$-channel DVCS, $\eta\rightarrow2\gamma$ and $\omega\rightarrow\pi^0\gamma$, in addition to the relative normalization of the expected 2$\gamma$ efficiency (from scenario 1) to extract the yield.}

\item{$e^\prime$ and 2$\gamma$ are detected. Although there are ongoing experimental efforts to ensure the detection of the forward recoiling proton, there is a small possibility the proton signal is rejected as background, which will complicate this scenario.  Here, the coplanarity of the two $\gamma$ that hit the ZDC will play a significant role in identifying $\pi^0$ events, and the reconstructed massing mass distribution may resolve the missing proton at the desired kinematics setting.} 

\end{itemize}

Background $\omega\rightarrow \pi^0\gamma$, the three gamma decay mode of $\omega$, has a branching ratio of 8.28\%~\cite{PDG2020}. Although it is possible for $\omega\rightarrow \pi^0\gamma$ to contaminate the $\pi^0$ event sample in all three trigger scenarios, it is possible to minimize this effect: examine angular coplanarity (back-to-back) in the center of mass frame for the two $\gamma$ that hit the ZDC; initiate a boundary in the missing mass ($m_{miss}$ or $m_{miss}^2$) distributions to exclude $\omega$ events. A full simulation study should give further insight on the level of experimental background and the most effective methodologies for removing them. 

\subsection{Electromagnetic Observables and TDAs at $\overline{\textrm{P}}$ANDA }
\label{sec:panda}

Baryonic exchange in antinucleon-nucleon collisions dominate both the small $-t$ and small $-u$ kinematical regions, i.e. both forward and backward angle physics. In fact, charge conjugation relates these two domains, and their theoretical description is in common while the experimental requirements are different. The electromagnetic processes, where a timelike virtual photon (of virtuality $q^2 > 0$ ) is produced in one of these regions, while a meson is created in the other region, are particularly well suited to study the predicted transition from a soft regime to a hard one as $q^2$ grows \cite{Pire:2005ax,Lansberg:2012ha}. The feasibility of measuring TDAs in proton anti-proton annihilations with $\overline{\textrm{P}}$ANDA at the future FAIR facility has been investigated in Refs. \cite{Singh:2014pfv} and \cite{Singh:2016qjg}. $\overline{\textrm{P}}$ANDA will use an anti-proton beam with a momentum of up to 15 GeV/$c$ and a luminosity of up to 2~$\times$10$^{32}$ cm$^{-2}$s$^{-1}$, interacting with a hydrogen target. So far, the feasibility of two reactions which can be used to extract TDAs has been investigated \cite{Singh:2014pfv,Singh:2016qjg}:
\begin{equation}
\overline{p} + p \rightarrow \gamma^* + \pi^0 \rightarrow e^+ + e^- + \pi^0 
\end{equation}
\begin{equation}
\overline{p} + p \rightarrow J/\psi + \pi^0 \rightarrow e^+ + e^- + \pi^0 
\end{equation}         

Within the factorized description in terms of $\pi N$-TDAs \cite{Pire:2005ax,Lansberg:2012ha}, the differential cross sections of these reactions are expected to show the following behavior:
\begin{equation}
\frac{d\sigma}{d\theta dq^{2}} \sim \frac{1}{(q^{2})^{5}} (1+\cos^{2}\theta),
\end{equation}
where $\theta$ is the polar angle of the lepton in the $e^{+}e^{-}$ CM frame in the case of the first reaction, or the polar emission angle of the $e^{+}$ or $e^{-}$ in the $J/\psi$ reference frame relative to the direction of motion of the $J/\psi$ in the case of the second reaction. Therefore, measuring the cross section in terms of $q^{2}$ can validate the characteristic scaling behavior predicted by the TDA model, while the $cos(\theta)$ dependence can be used to check the predicted dominance of the transverse polarization of the virtual photon. In addition, the measurements with $\overline{\textrm{P}}$ANDA, in combination with the results from electroproduction studies, will provide evidence for the universality of the TDA model \cite{Singh:2014pfv,Singh:2016qjg}.

For both channels, detailed Monte Carlo simulations, based on a realistic TDA-model-based event generator and the rejection of several background channels, has been investigated for different center of mass energies. It has been found that a sufficient background rejection can be achieved to provide a clean measurement of both channels \cite{Singh:2014pfv,Singh:2016qjg}.
For $\overline{p} + p \rightarrow \gamma^* + \pi^0$, the $q^{2}$ and $\cos(\theta)$ dependencies can be well measured with an integrated luminosity of 2~fb$^{-1}$, which corresponds to approximately half a year of data taking \cite{Singh:2014pfv}. With this integrated luminosity, a statistical uncertainty of the cross section in the order of 12 - 24 \%, depending on the CM energy, can be achieved, which is sufficient to validate the model predictions \cite{Singh:2014pfv}. For $\overline{p} + p \rightarrow J/\psi + \pi^0$, the $\cos(\theta)$ dependence of the cross section can be measured with a statistical uncertainty $\Delta\sigma(t,u) /\sigma(t,u) \sim$ 5 - 10 \% for an integrated luminosity of 2~fb$^{-1}$, which is sufficient to validate the predicted behavior \cite{Singh:2016qjg}.
With higher integrated luminosities, $\overline{\textrm{P}}$ANDA will enable detailed multidimensional measurements of the cross section and can contribute to the extraction of TDAs in global analyses.

\section{Summary and Conclusions}

The {\em Backward angle (u-channel) Physics Workshop} put into perspective a new domain of hadronic physics, namely the transition from soft baryon-exchange processes (where the scattering amplitude is described in terms of baryon or Reggeized baryon trajectories), to a partonic description in a deep regime (where a large scale allows one to resolve the quark and gluon constituents of hadrons). This transition is expected in a specific kinematic domain where the existence of a short distance process ensures the factorization of scattering amplitudes into perturbatively calculable hard subprocess amplitudes and well-defined matrix elements of operators on the light-cone. 
More checks are clearly needed before this transition clearly appears, and 
both experimental and theoretical progresses are needed. 
On the theoretical side, it includes
\begin{itemize}
    \item 
improving the Regge description of various processes with a virtual photon, including polarization observables, which have often been decisive to test different parametrizations of necessary phenomenological inputs.

\item improving the QCD framework to obtain information on partonic quantities through non-perturbative techniques such as lattice QCD or light cone QCD sum-rules and estimating  higher twist corrections to amplitudes.

\end{itemize}
On the experimental side, much progress is expected from
\begin{itemize}
    \item  the photoproduction program (both in UPC reactions and at JLab-GlueX), where backward timelike Compton scattering or heavy meson production may help to scan the transition to the partonic regime.

\item 
the electroproduction program at JLab (both CLAS12 and Hall~C), where backward production of various mesons, as well as backward DVCS, should provide further information on the onset of the partonic domain first observed in $\pi$ and $\omega$ electroproduction data, and allow one to study the possible universality of this transition.

\item the electromagnetic processes initiated in antiproton-nucleon annihilation at $\overline{\textrm{P}}$ANDA with the production of a lepton pair or a heavy meson accompanied by various mesons.

\item similar processes initiated in meson-nucleon collisions (at J-PARC and COMPASS at CERN) in kinematics where a baryonic exchange dominates.

\item the use of nuclear targets to develop a nuclear transparency program testing the color transparency property of partonic reactions.

\item  opportunities presented by high energy electron-ion colliders, such as the EIC, which will extend the $u$-channel electroproduction program into the high energy, high $Q^2$ regime and provide data on a wide range of $u$-channel reactions, including for heavy (charmonium) mesons.
\end{itemize}
Such measurements will hopefully allow one to answer the questions highlighted throughout this workshop: do we observe a transition from a soft to a hard regime in the processes with a baryon charge exchange, and which new information on the nucleon structure do we extract from them?

In conclusion, backward angle ($u$-channel) production exhibits many surprising features. While much pioneering experimental and theoretical work has been done to elucidate and understand these observations, it is clear that more research is required. With the currently proposed backward angle measurements, and the resulting theoretical analyses, we expect that the next decade, and beyond, should bring further progress in understanding this process.


\section{Acknowledgements}

We thank the Jefferson Science Associates (JSA) Initiatives Fund Program and the Jefferson Lab for sponsoring the {\em Backward angle (u-channel) Physics Workshop.}

RJP was supported by Taiwanese MoST Grant No. 109-2112-M-009-006-MY3 and Taiwanese MoST Grant No.
109-2811-M-009-516.

SK and AS were supported by the U.S. Department of Energy, Office of Science, Office of Nuclear Physics, under contract number DE- AC02-05CH11231.

JRS and WBL are supported by the U.S. Department of Energy, Office of Science, Early Career Award contract DE-SC0018224.

WBL was supported as a postdoctoral fellow from the Jefferson Lab Electron-Ion Collider Center.
WBL also acknowledges the financial support from Jefferson Science Associates through the Post-Doctoral Award.

GMH and SJDK are supported by the Natural Sciences and Engineering Research Council of Canada (NSERC), SAPIN-2021-00026.

LS is supported by grant 2019/33/B/ST2/02588 of the National Science Center in Poland.

B-G Yu was supported by the National Research Foundation of
Korea grant NRF-2017R1A2B4010117.

{\L}B acknowledges the financial support and hospitality of the Theory Center at  Jefferson Lab and Indiana University.
{\L}B was also supported by the
Polish Science Center (NCN) grant 2018/29/B/ST2/02576.

VM is a Serra Húnter fellow and acknowledges support from the Spanish national Grant No. PID2019–106080 GB-C21 and PID2020-118758GB-I00.

The work of KS is supported
by the Foundation for the Advancement of Theoretical Physics and Mathematics ``BASIS''.

 This project is also co-financed by the Polish-French collaboration agreements Polonium, by the Polish National Agency for Academic Exchange and COPIN-IN2P3 and by the European Union’s Horizon 2020 research and innovation program under grant agreement 824093.

\appendix

\section{Missing mass reconstruction technique}
\label{app:missmass}

\begin{figure}[htb]
\centering
\includegraphics[scale=0.40]{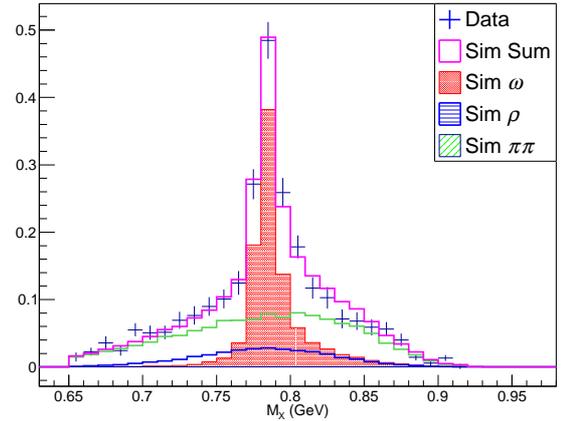}
\caption{Example reconstructed missing mass ($m_{miss}$) distribution for $ep\rightarrow e^{\prime} p^{\prime} X$ at $Q^2 = 2.45$~GeV$^2$ (blue points) from Hall~C of JLab \cite{Li:2019xyp,wenliang17}. The simulated distributions for $\rho$ (blue), $\omega$ (red) and $\pi\pi$ (green) are used for the reaction channel separation.}
\label{fig:missmass}
\end{figure}

The missing mass $m_{miss}$ for a meson $X$ ($X = \pi_0$, $\omega$, etc.) production interaction $^1$H$(e, e^{\prime}p)X$:
\begin{equation}
e\,(p_e) + p\,(p_p) \rightarrow e\,(p_{e^{\prime}})+ p\,(p_{p^\prime}) + X \,.
\end{equation}
is calculated as:
\begin{equation}
\begin{aligned}
\hspace{4em}  m_{miss} = 
    \big\{ 
    &\left(E_{e} + m_{p} - E_{e'} - E_{p'}\right)^{2} - \\
    &\left(\vec{p}_{e} - \vec{p}_{e'} - \vec{p}_{p'}\right)^{2}
    \big\} ^{1/2}
\end{aligned}
 \label{eqn:missmass}
\end{equation}
For instance, in the case of  $\omega$ electroproduction,  the $\omega$ events sit on a broad background, shown in reconstructed missing mass spectrum for $ep\rightarrow e^{\prime} p X$ in Fig.~\ref{fig:missmass}. The final state particle $X$ could be: $\omega$, $\rho$ or non-resonant $\pi\pi$. Various quality control criteria were introduced to validate the background subtraction procedure, as described in Ref.~\cite{wenliang17}.

\bibliographystyle{epj}

\bibliography{refs}

\end{document}